\RequirePackage{fix-cm}
\documentclass[smallextended]{svjour3}       
\smartqed  
\usepackage{graphicx}
\usepackage{color}
\begin{document}
\def\tmttfbr{$\mathrm{(TMTTF)_{2}Br}$\,}
\def\PF{$\mathrm{(TMTSF)_{2}PF_{6}}$\,}
\def\TM{(TM)$_2$X}
\def\pc{$P_{c}$\,}
\def\hc2{$H_{c2}$\,}
\def\tc{$T_{c}$\,}
\def\tm6{$\mathrm{(TMTSF)_{2}PF_{6}}$\,}
\def\t6as{$\mathrm{(TMTSF)_{2}AsF_{6}}$}
\def\tmx{$\mathrm{(TMTSF)_{2}(ClO_{4})_{(1-x)}ReO_{4}_x}$\,}
\def\tmc{$\mathrm{(TMTSF)_{2}ClO_{4}}$\,}
\def\tms{$\mathrm{(TMTSF)_{2}AsF_{6(1-x)}SbF_{6x}}$}\,
\def\tmps{$\mathrm{(TMTTF)_{2}PF_{6}}$\,}
\def\tmttfsbf6{$\mathrm{(TMTTF)_{2}SbF_{6}}$\,}
\def\tmttfasf6{$\mathrm{(TMTTF)_{2}AsF_{6}}$\,}
\def\tmttfbf4{$\mathrm{(TMTTF)_{2}BF_{4}}$\,}
\def\tmtsfreo4{$\mathrm{(TMTSF)_{2}ReO_{4}}$\,}
\def\tm2x{$\mathrm{(TM)_{2}X}$\,}
\def\tq{$\mathrm{TTF-TCNQ}$\,}
\def\tsq{$\mathrm{TSeF-TCNQ}$}\,
\def\qnq{$(Qn)TCNQ_{2}$}\,
\def\R{$\mathrm{ReO_{4}^{-}}$}  
\def\C{$\mathrm{ClO_{4}^{-}}$}
\def\P{$\mathrm{PF_{6}^{-}}$}
\def\tqr{$\mathrm{TCNQ^\frac{\cdot}{}}$\,}
\def\nmpq{$\mathrm{NMP^{+}(TCNQ)^\frac{\cdot}{}}$\,}
\def\q{$\mathrm{TCNQ}$\,}
\def\nmp{$\mathrm{NMP^{+}}$\,}
\def\f{$\mathrm{TTF}\,$}
\def\tc{$T_{c}$\,}
\def\nmq{$\mathrm{(NMP-TCNQ)}$\,}
\def\ts{$\mathrm{TSF}$}
\def\tsm{$\mathrm{TMTSF}$\,}
\def\tst{$\mathrm{TMTTF}$\,}
\def\tmp6{$\mathrm{(TMTSF)_{2}PF_{6}}$\,}
\def\tms2x{$\mathrm{(TMTSF)_{2}X}$}
\def\tm2x{$\mathrm{(TM)_{2}X}$\,}
\def\as{$\mathrm{AsF_{6}}$}
\def\sb{$\mathrm{SbF_{6}}$}
\def\pf{$\mathrm{PF_{6}}$}
\def\re{$\mathrm{ReO_{4}}$}
\def\ta{$\mathrm{TaF_{6}}$}
\def\cl{$\mathrm{ClO_{4}}$}
\def\4fb{$\mathrm{BF_{4}}$}
\def\ttdm{$\mathrm{(TTDM-TTF)_{2}Au(mnt)_{2}}$}
\def\edt{$\mathrm{(EDT-TTF-CONMe_{2})_{2}AsF_{6}}$}
\def\tfx{$\mathrm{(TMTTF)_{2}X}$\,}
\def\tsx{$\mathrm{(TMTSF)_{2}X}$\,}
\def\tmx{$\mathrm{(TMTSF)_{2}(ClO_{4})_{(1-x)}(ReO_{4})_{x}}$\,}
\def\ttftcnq{$\mathrm{TTF-TCNQ}$\,}
\def\ttf{$\mathrm{TTF}$\,}
\def\tcnq{$\mathrm{TCNQ}$\,}
\def\bedtttf{$\mathrm{BEDT-TTF}$\,}
\def\reo4{$\mathrm{ReO_{4}}$}
\def\bedtttfreo4{$\mathrm{(BEDT-TTF)_{2}ReO_{4}}$\,}
\def\et2i3{$\mathrm{(ET)_{2}I_{3}}$\,}
\def\et2x{$\mathrm{(ET)_{2}X}$\,}
\def\ket2x{$\mathrm{\kappa-(ET)_{2}X}$\,}
\def\cuncnbr{$\mathrm{Cu(N(CN)_{2})Br}$\,}
\def\ket2x{$\mathrm{\kappa-(ET)_{2}X}$\,}
\def\cuncncl{$\mathrm{Cu(N(CN)_{2})Cl}$\,}
\def\cuncs{$\mathrm{Cu(NCS)_{2}}$\,}
\def\betsfecl4{$\mathrm{(BETS)_{2}FeCl_{4}}$\,}
\def\bets{$\mathrm{BETS}$\,}

\def\et{$\mathrm{ET}$\,}

\title{Organic Superconductors: when correlations and magnetism walk in
}

\author{Denis J\'erome       
}


\institute{D. J\'erome \at
               Laboratoire de Physique des Solides, Universit\'e Paris-Sud, 91405 Orsay, France \\
             Tel.: +(33)169155301\\
              \email{denis.jerome@u-psud.f}           
}

\date{Received: date / Accepted: date}

\maketitle

\begin{abstract}
This  survey   provides a brief account for the start of organic superconductivity motivated by the quest for high \tc superconductors and its  development since the eighties'.  Besides superconductivity found in 1D organics in 1980, progresses in this  field of research have contributed to  better understand the physics of low dimensional conductors  highlighted  by the wealth of new remarkable properties. Correlations conspire to govern the low temperature properties of the metallic   phase.  The contribution of antiferromagnetic fluctuations  to   the interchain  Cooper pairing proposed by the theory is borne out by  experimental investigations and supports supercondutivity  emerging from a non Fermi liquid background.  Quasi one dimensional organic superconductors can therefore be considered as simple prototype systems for the more complex high \tc materials.

\end{abstract}


\section{Superconductivity in the seventies}
\label{sec:1}

Bardeen, Cooper and Schrieffer (BCS)\cite{Bardeen57}  had  emphasized in 1957 the existence of a two body   attractive interaction between charge carriers (either electrons or holes)
 being a prerequisite for a Bose condensation of electron pairs into the superconducting state.
 This net attractive coupling in spite of the Coulomb repulsion between carriers of the same sign relies closely on an attraction between electrons mediated by their interaction with excitations of the lattice namely, the phonons. 
 In addition, a major achievement of the BCS theory has been the understanding  of the ionic mass dependence of the critical temperature i.e. (\tc $\alpha \,M^{-1/2}$).
Although Fr\"ohlich had proposed  in 1954 a model for superconductivity based on the involvement of the lattice  \cite{Frohlich54}
this theory was  unable to account for the superconductivity of metals but turned out to be  quite 
relevant later for the interpretation of the transport properties in some one dimensional organic compounds, \textit{vide infra}. 
\begin{figure*}[htbp]	
\centerline{ \includegraphics[width=0.9\hsize]{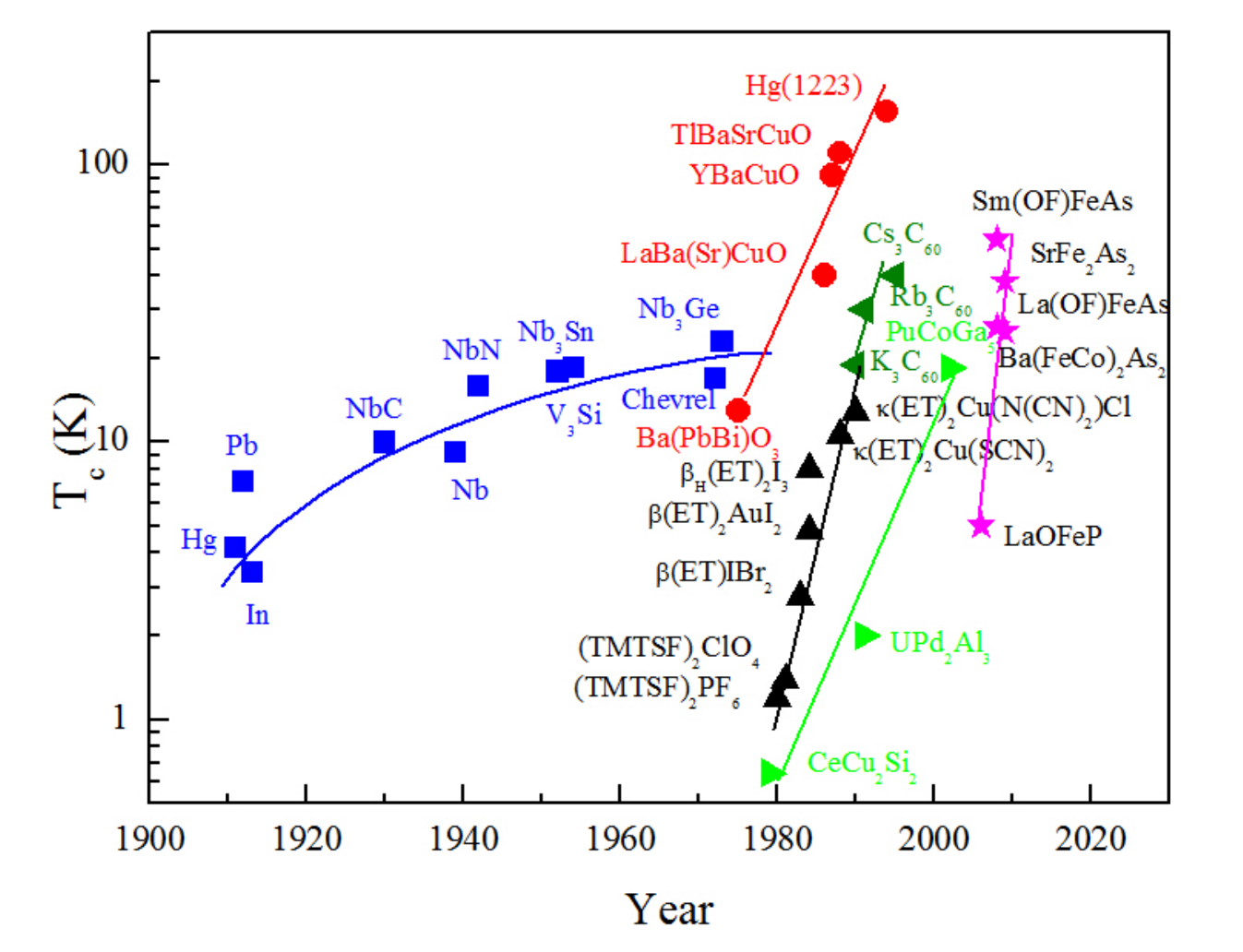}}
\caption{Evolution of \tc in various series of  superconductors over the years. }
\label{Tc-Year-2010.pdf} 
\end{figure*}

The search for new materials exhibiting increased critical temperatures \tc has been a strong motivation for research since the initial discovery of superconductivity  by Kamerlingh Onnes\cite{Onnes11} in 1911. The rise of \tc has been relatively slow during the first 50 years as shown on fig(\ref{Tc-Year-2010.pdf}) with some kind of a saturation in the sixties.

At the beginning of the 60's  
 the term \textit{high temperature superconductor} was already commonly used refering
to the intermetallic compounds of the A15 structure, namely, ($\mathrm{Nb_{3}Sn}$ or $\mathrm{V_{3}Si}$)\cite{Hardy54} see fig(\ref{Tc-Year-2010.pdf}). These superconductors  are still  until now  the materials used for building most superconducting magnets.

 However, some new concepts for superconductivity had been recognized by Professor Friedel namely, the hidden 1D nature of the A15 structure which provides an enhancement of the density of states at the Fermi level
lying close to the van-Hove singularity of the density of states. Within the BCS formalism  large \tc could in turn
be expected. They were actually observed (17-23 K) but an upper limit was found to the increase of \tc since the  large value of
N(E$_{\mathrm{F}}$) makes also the  structure unstable against a  cubic to tetragonal band Jahn-Teller distortion\cite{Labbe66,Weger73}.
 Labb\'e, Bari\v si\'c and Friedel made the remark in 1967  that \tc's of A15 compounds were not much influenced by the characteristic energy of the phonon spectrum  but determined  by the narrowness of the density  of states at Fermi level. They also showed   that the band filling of actual compounds such as 
$\mathrm{Nb_{3}Sn}$ and $\mathrm{V_{3}Si}$ makes their \tc's optimum\cite{Labbe67}. Consequently, based on metallurgical considerations,
some authors at the end of the seventies  considered 25-30 K,  as an upper limit   for
\tc\cite{Hulm80}. 
\subsection{Some explorers}

Although, at that time all attempts to increase \tc were still based on the BCS phonon-mediated theory with its  strong coupling extension\cite{Migdal58,Eliashberg60} some new paths were proposed.

A few years after the BCS paper, Kohn and Luttinger proposed a new mechanism for superconductivity\cite{KohnLuttinger65}  still based on the pairing idea although  the attraction is no longer phonon-mediated. The attraction derives from an extension of Friedel'density oscillations around  charged impurities in metals\cite{Friedel54,Friedel58}. This mechanism is an entirely electronic process. However, the expected critical temperature should be extremely small  in the milli Kelvin range but Kohn and Luttinger emphasized that flat Fermi  surfaces or van Hove singularities could greatly enhance the actual \tc.

Expending the successful idea of the isotope effect in the BCS theory  other models were proposed in which excitations of the lattice responsible for the electron pairing had been replaced by higher energy excitations namely, electronic excitations, with the hope of  new  materials with  \tc higher than those explained by the BCS theory.
Consequently, the small electronic mass
$m_e$ of the polarizable medium would lead  to an enhancement of
\tc of the order of  ($M/m_e)^{1/2}$ times the value which is observed  in a conventional superconductor, admittedly a huge factor. V.L.Ginzburg\cite{Ginzburg64a,Ginzburg64b} considered in 1964 the possibility for the pairing of electrons in
metal layers sandwiched between polarizable dielectrics through virtual excitations at high energy. 

It is illuminating to have a look on fig(\ref{Tc-Year-2010.pdf}) at the compound $\mathrm{Ba(Pb_{0.8}Bi_{0.2})O_3}$ reported  to become superconducting at 13 K\cite{Sleight75}. This was in 1975 the highest \tc ever observed for an oxide. With this high temperature superconductivity in the perovskite series  $\mathrm{Ba(PbBi)O_3}$,  \textit{D$\acute{e}j\grave{a}\,  Napol\acute{e}on\, pointait\, sous\, Bonaparte!$} But, as Edelsack, Gubser and Wolf wrote in their introduction for the conference proceeding on Novel Superconductivity edited by Wolf and Kresin\cite{Edelsack87}, \textit{Organics were those who decided to forge revolutionary paths in the quest for high} \tc.

In this respect, W.A.Little had
made an interesting suggestion in 1964\cite{Little64,Little65}:
 a new mechanism  for superconductivity supposed to provide to a drastic enhancement of the superconducting \tc
 to be observed in especially designed macromolecules. The idea of Little
was indeed strongly rooted in the extension of the isotope effect proposed by BCS.

However, a prerequisite to the model of Little was the achievement of conduction in molecular crystals namely, new types of conductors in which the conduction would proceed through
the transport of charge between molecular orbitals rather than atomic orbitals. 

Actually, the concept of a synthetic metal had
aleady been  launched   by McCoy and Moore\cite{McCoy11}  when they
proposed:
\textit{to prepare composite metallic substances from non-metallic constituent elements}. 
 As to the possibility of superconductivity in materials other than metals, Fritz London in 1937\cite{London37} has been the first to
suggest that  compounds with aromatic rings such as anthracene, naphtalene,... \textit{might exhibit a
 current running freely around the rings under a magnetic field}. 
 
 The first successful  attempt  to
promote metal-like conduction between open shell molecular species came out in 1954 with the synthesis of the molecular salt of
perylene oxidized with bromine \cite{Akamatsu54} although this salt was rather unstable.
\subsection{Organic metallicity}
Coming back to 1D materials, the model of Little had not much in common with the early  remark of London. It  was based on the use of a long conjugated polymer such as a
polyene molecule grafted by polarizable side groups\cite{Little70}. Admittedly, this formidable
task in synthetic chemistry has not reached its initial goal but the idea to link organic metallicity and one dimensionality  was
launched and turned out to be a very strong stimulant for the development of organic superconductors. 

One dimensional conductors had been regarded previously   as  textbook examples by
Peierls\cite{Peierls55}, who stated a theorem according to which a gap should spontaneously open at the Fermi level in the conduction 
and lead to the stabilization of a dielectric insulating state at low temperature ruining therefore any further hope for superconductivity. The  point is that the seminal paper of Little\cite{Little64} which was still based on the popular Fermi liquid BCS
theory had made the assumption that the enhanced Cooper pair attraction would be strong enough to overcome the energy gain of a Peierls-distorted chain.
Needless to say that  a lot of
basic  problems namely, the possibility of synthesizing conducting polymer chains and the competition against a Peierls distorsion, had been overlooked.   
Eighteen years later the Peierls transition   has been established in 
platino-cyanate chains\cite{Comes73}.   

Bychkov \textit{et-al}\cite{Bychkov66}  criticized  the model of Little regarding its potentiality   to
lead to high temperature superconductivity. Following Bychkov \textit{et-al}, the one-dimensional character of the model system proposed by Little
makes it a unique problem in which there exists a  built-in coupling between superconducting and dielectric instabilities.  It follows
that each of these instabilities cannot be considered separately in the mechanism proposed by Little in one dimension and last but not
least, fluctuations should be very efficient in a 1D conductor to suppress any long range ordering to very low temperature. These remarks proved to be very pertinent for the development several decades later of the adequate theoretical framework needed for organic superconductivity.

A major step was accomplished in 1970 towards the discovery of new materials for superconductivity with the synthesis by F.Wudl\cite{Wudl70} of the new molecule tetrathiafulvalene, (\ttf).
This molecule 
containing four sulfur  heteroatoms in the fulvalene skeleton can easily donate electrons when it is combined to electron accepting species.
This discovery  has deeply influenced the subsequent evolution of the chemistry of organic conductors. 

\subsection{The  charge transfer period, first organic conductors}
\label{sec:1.1}
The \ttf molecule 
containing four sulfur  heteroatoms in the fulvalene skeleton can easily donate electrons when it is combined to electron accepting
molecules allowing the synthesis of the first stable organic metal,the charge transfer complex, \ttftcnq. The system is made up of two
kinds of flat molecules each forming  segregated parallel conducting stacks. This compound can   be recognized as an organic conductor
as the orbitals involved in the conduction ($\pi$-HOMO and $\pi$-LUMO for \ttf and
\tcnq respectively) are associated with the molecule as a whole rather than with a particular atom with carriers in each stacks
 provided by an  interstack charge  transfer \textit{at variance} with other organic conductors such as the doped conjugated polymers.

The announcement of a large and metal-like conduction in \ttftcnq \\was made in 1973, simultaneously  by the Baltimore\cite{Ferraris73}
and  Pennsylvania\cite{Coleman73} groups.  The Pennsylvania group made a provocating claim announcing  a giant conductivity peak of the
order of
$10^5$
$(\Omega cm)^{-1}$ at 60 K arising just above a very sharp transition towards an insulating ground state at low temperature. This
conductivity peak was attributed by their authors to precursor signs  of an incipient superconductor.  Unfortunately,
the  conductivity peak with such a giant amplitude could never be reproduced by other groups who anyway all agreed on  the metallic
character of this novel molecular material\cite{Thomas76}. Besides,  the Orsay group showed  from X-ray scattering studies that the metal
insulator transition at 59 K was the consequence of the instability of a conducting chain predicted by Peierls\cite{Denoyer75}. The X-ray diffuse scattering study did provide an important information namely, the band filling of \tq is incommensurate with a charge transfer of $\rho=0.59$ at 300 K\cite{Denoyer75}. 

\begin{figure*}[htbp]	
 \centerline{\includegraphics[width=0.5\hsize]{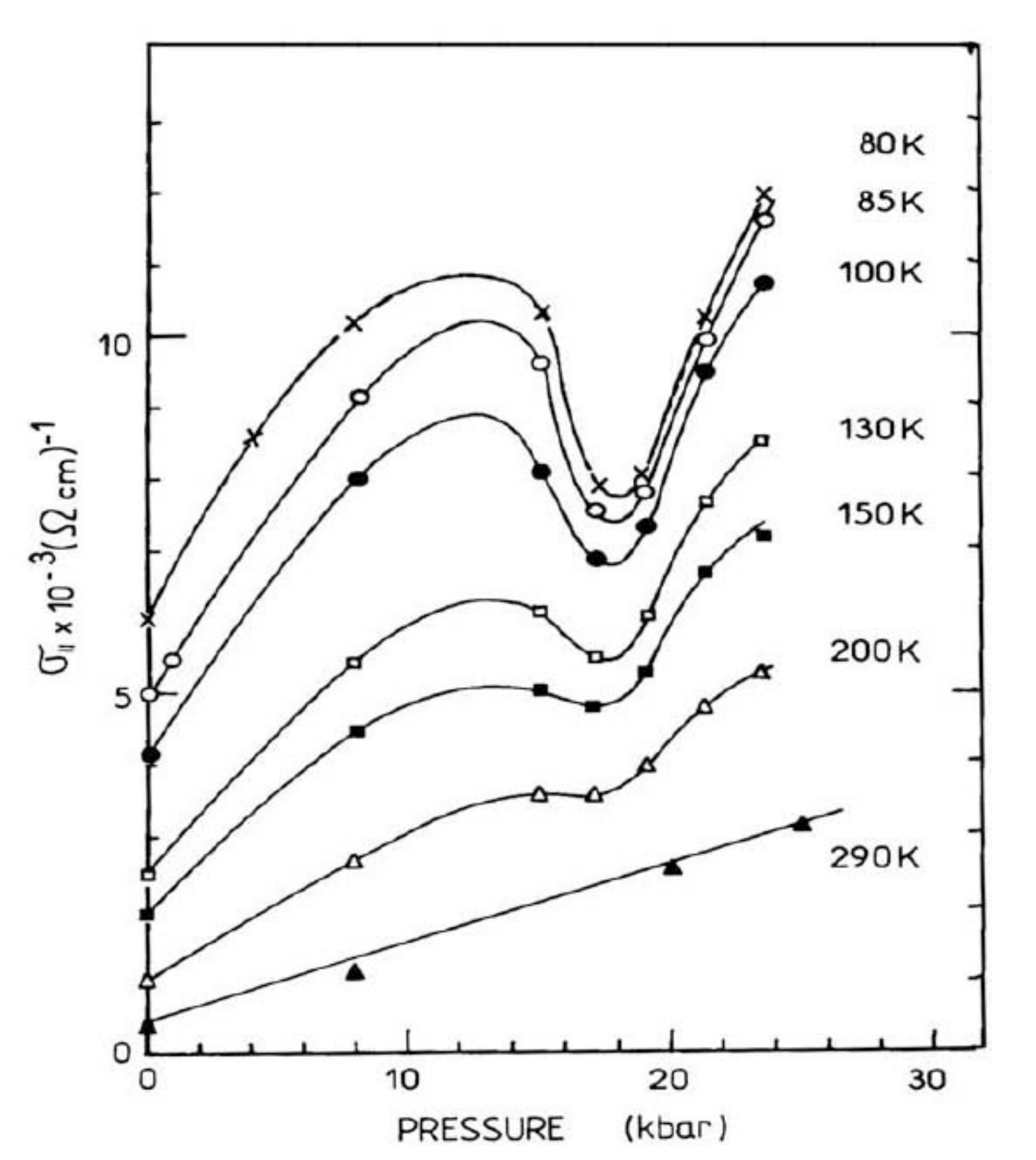}}
\caption{Isothermal pressure dependence of the longitudinal conduction of \tq \,  showing a drop of the conduction in the commensurability (x3) domain  of the fluctuating Fr\"ohlich contribution due to pinning by the lattice\cite{Jerome82}.  The ratio $\sigma (80 K)/\sigma (290 K)$ is evolving from 18 to 3 from ambient pressure up to 18 kbar and rising again at higher pressures.   }
\label{TQfluctuations.pdf} 
\end{figure*}

It is  the
study of transport properties under pressure which settled the origin of the conductivity peak of the conducting phase  before the Peierls transitioin at 59. Driving the charge transfer through a commensurate value $\rho= 2/3$ under the pressure of $\approx$ 18 kbar provided the proof for collective  Fr\"ohlich fluctuations in a 1D regime leading in turn to a large enhancement  of the  ordinary single particle conduction as long as the wave length of these $CDW$ fluctuations  is not commensurate with the underlying lattice \textit{in particular} under ambient pressure\cite{Friend78}, see fig(\ref{TQfluctuations.pdf}). 
 High  pressure appeared to be 
a  parameter far more influencial for organic conductors than for regular metals, it  suppressed the onset  of the  Peierls instability in \ttftcnq\cite{Yasuzuka07} but failed to reveal any sign of superconductivity possibly with a high \tc   as expected from the work of Horovitz \textit{et-al}\cite{Horovitz75} when phonons become soft in the vicinity of the Peierls transition.

\begin{figure*}[htbp]	
 \centerline{\includegraphics[width=0.6\hsize]{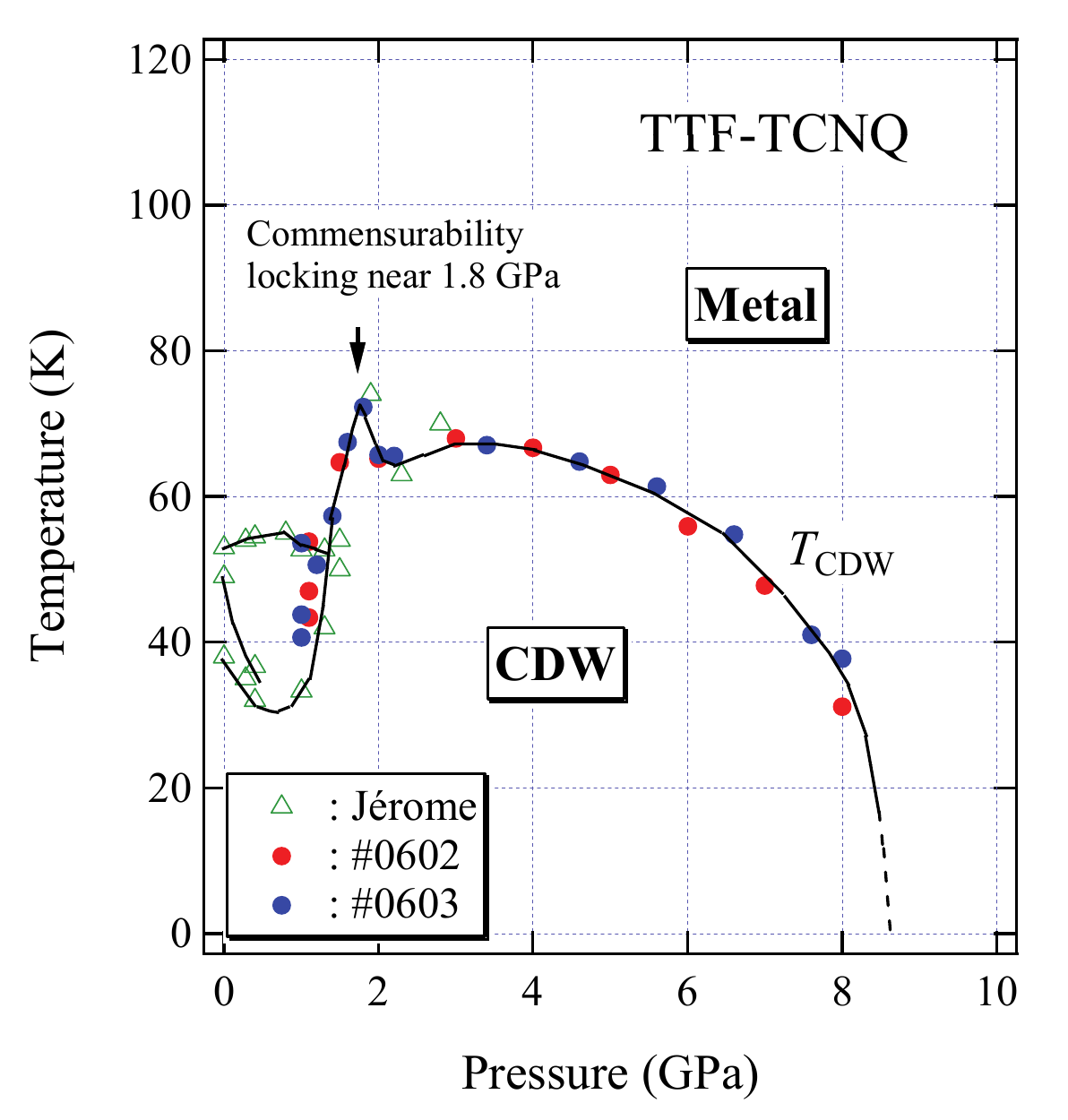}}
\caption{Phase diagram of \tq under pressure measured by the Osaka group\cite{Yasuzuka07} including the  early  Orsay data\cite{Friend78} up to 30 kbar. The position of the peak at $\approx$ 18 kbar for the Peierls transition corresponds to the dip of longitudinal conduction on fig(\ref{TQfluctuations.pdf}).}
\label{DiagTTFTCNQ.pdf} 
\end{figure*}

The Orsay high pressure experiments on \tq have been  extended recently up to the pressure domain of 80 kbar by the Osaka group\cite{Yasuzuka07} as shown on fig(\ref{DiagTTFTCNQ.pdf}). The early high pressure Orsay results  up to 30 kbar with commensurability arising around 18 kbar have be confirmed but in addition the Peierls transition is suppressed down to 31 K at 80 kbar, probably an effect of increased interstacks coupling.

Furthermore, the role of 1D electron-electron repulsive interactions
has been recognized by the $4k_F$ signature in X-ray diffuse scattering experiments\cite{Pouget76}. 
Since decreasing the size of the Coulombic repulsion is expected to boost the conductivity of metals,  other
synthetic routes have then been followed. In the 1970s, the leading ideas governing the search for new materials likely to exhibit good
metallicity and possibly superconductivity were driven by the possibility to minimize the role of electron-electron repulsions and at the
same time to increase the electron-phonon interaction while keeping the overlap between stacks as large as possible. This led to the
synthesis of new series of charge transfer compounds which went beyond the known \tq system, for example changing the
molecular properties while retaining the same crystal structure. It was recognized that the electron polarizability is important
to reduce the screened on-site electron-electron repulsion and that the redox potential $(\Delta E)_{1/2}$ should be minimized,
\cite{Garito74,Engler77} in order to fulfill this goal. Hence, new charge transfer compounds with the acceptor \q  were synthesized using
other heteroatoms for the donor molecule, 
\textit{i.e.} substituting sulfur for selenium in the \f skeleton thus leading to the \ts \, molecule where S stands for selenium Se. 

Much attention was put on the
tetramethylated derivative of the \ts \, molecule which, when combined to the dimethylated \q gave rise to TMTSF-DMTCNQ\cite{Andersen78}.The
outcome of the high pressure study of this latter compound has been truly  decisive for the quest of organic superconductivity
\cite{Andersen78,Jacobsen78,Tomkiewicz78}. First, the metal insulator transition located at 42 K has been identified by  X-ray diffuse scattering experiments\cite{Pouget81} as a Peierls transition driven by the TMTSF chain and a quarter filling of both donor and acceptor bands of this charge transfer compound has been derived   from the measurement of the wave vector $2k_F$ in the Peierls state. Second, for the first time the metallic state of an organic compound could be stabilized down to 1.2 K  under pressure    \textit{albeit} without superconductivity \cite{Andrieux79a}. Even if the reason for the stability of a metallic state in  TMTSF-TCNQ  above the pressure of 9 kbar is not fully established at the moment emphasizing the role of the donor stack (TMTSF)  has been a strong motivation for the synthesis of new organic conductors with  structures even simpler than those of   two stacks charge transfer compounds. Fortunately, the Montpellier group\cite{Galigne78} had synthesized in 1978 a series
of isomorphous radical cationic conductors based on TMTTF (the sulfur analog of the TMTSF molecule) with an inorganic
anion namely, \tfx. These materials were all insulating at ambient pressure although  some of them   did reveal superconductivity under
pressure much later and became quite important for studying the physics of 1D conductors since it is the replacement of TMTTF by TMTSF in the stoichiometric \TM\, salt which opened the door to organic superconductivity\cite{Bechgaard79}.
\section{A breakthrough in the eighties}
At the same time, the Copenhagen group led by Klaus Bechgaard and experienced with the chemistry of selenium  succeeded in the
synthesis of a new series of conducting salts all based on the  TMTSF  molecule  namely, \tms2x   where X is an inorganic
mono-anion with various possible symetry, spherical (\pf ,\as , \sb ,\ta  ), tetrahedral (\4fb ,\cl ,\,\re )
 or triangular $\mathrm{(NO_3}$)\cite{Bechgaard79}. All these compounds but the one with X= \cl \, did reveal an insulating ground state
under ambient pressure.  
\begin{figure*}[htbp]	
 \centerline{\includegraphics[width=1\hsize]{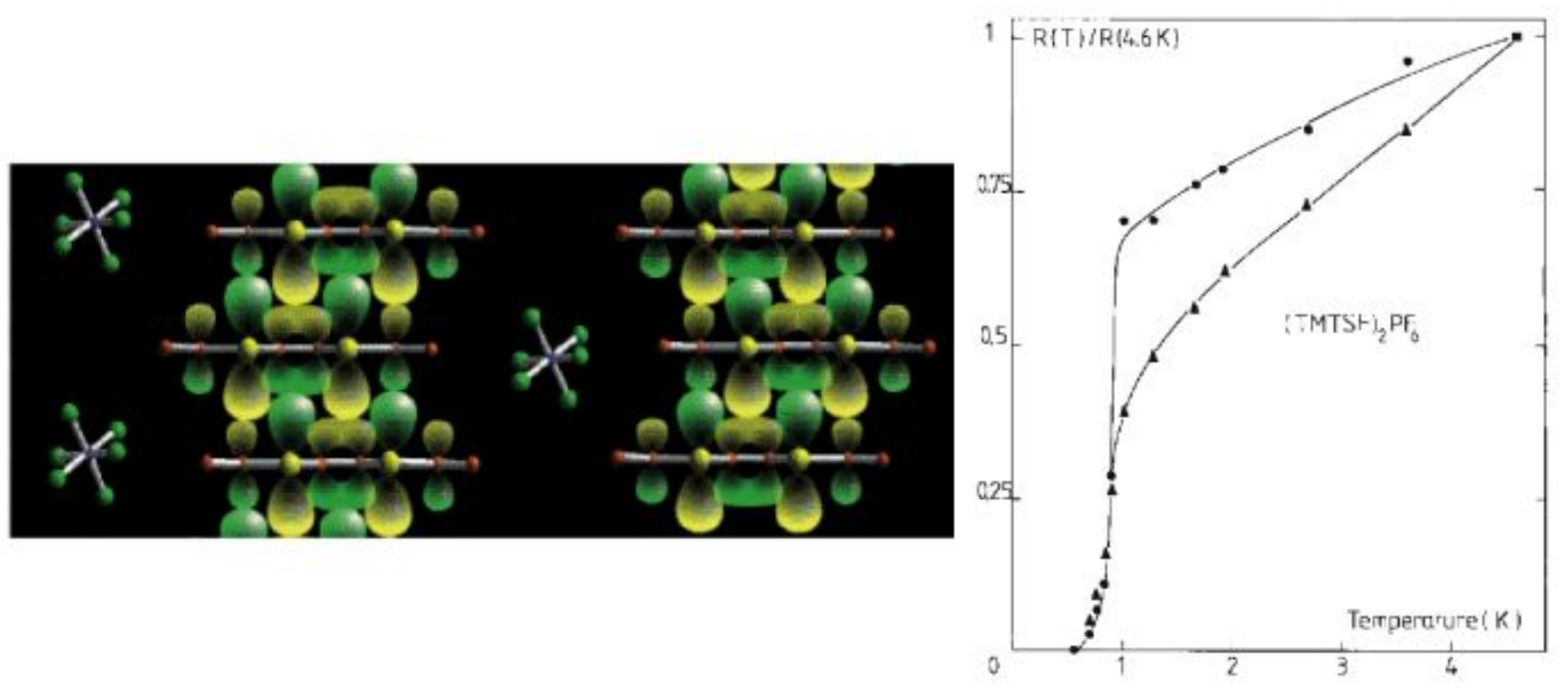}}
\caption{Side view of \tmc along the $b$ axis. The structure of \tmp6 is similar with  centrosymmetric \P anions.   First observation of superconductivity in \tmp6 under a pressure of 9 kbar\cite{Jerome80}. The resistance of two samples is normalized to the 4.5 K value. Two features are abnormal for the resistance of ordinary metals at low temperature, the strong temperature dependence below 4 K and the sublinear behaviour above \tc. They are both extensively discussed in Secs (3.4.1 and 3.4.3).}
\label{PF6supra.pdf} 
\end{figure*}
What is  so special with \tmp6 , the prototype of the so-called  Bechgaard salts, unlike previously investigated \tq ,  is the magnetic origin of the ambient pressure insulating state\cite{Andrieux81} contrasting with the Peierls-like ground states discovered until then. The ground state of \tmp6 turned out to be a spin density wave state similar to the predictions of Lomer\cite{Lomer62} and Overhauser\cite{Overhauser60a} for metals.  The onset of itinerant antiferromagnetism opens a gap at Fermi level and since the  Fermi surface is nearly planar the gap develops over the entire surface.  The magnetic origin of the insulating ground state of  \tmp6 was thus the first experimental hint for the prominent role played by correlations in these organic conductors, fig(\ref{PF6supra.pdf}). Superconductivity of \tmp6 occured at 1 K once the $SDW$ insulating ground state could be suppressed under a pressure of about 9 kbar\cite{Jerome80}. The competition between a distorted phase and  superconductivity  had already been observed for instance in the transition metal dichalcogenides layer crystals under pressure such as $\mathrm{2H-NbSe_2}$\cite{Molinie74}. However, in this latter situation superconductivity preexisted in the low pressure metallic phase with  a Peierls distortion due to the nesting of some regions of the Fermi surface\cite{Wilson75}decreasing the  density of states at the Fermi level. As noticed by Friedel\cite{Friedel74} the effect of pressure on the \tc of $\mathrm{2H-NbSe_2}$ compounds is likely to be an increase of $N(E_F)$ by reducing the gaps on some parts on the Fermi surface leading to a concomitant enhancement of of \tc. It is a situation \textit{at variance} with what is encountered in \tmp6 since the $SDW$ phase is truly insulating. While,  superconductivity is optimized at the border with the $SDW$ state, the reasons are quite different from what has been suggested for layer compounds as discussed below.
\subsection{A generic phase diagram}
\label{sec:2.1}
\begin{figure*}[htbp]	
 \centerline{\includegraphics[width=1.1\hsize]{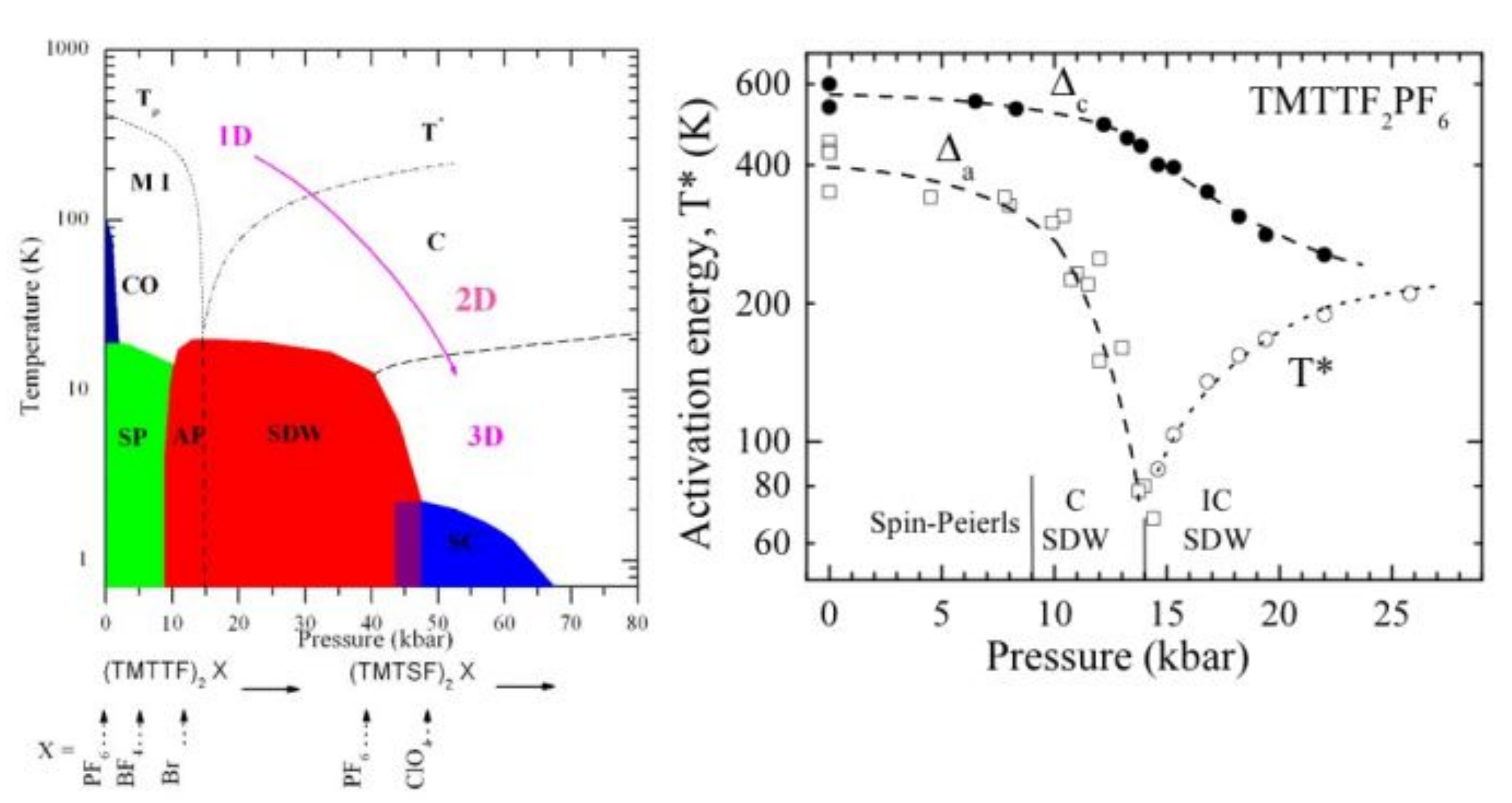}}
\caption{Left. Generic phase diagram for  \tm2x compounds\cite{Jerome91} based on the sulfur compound \tmps under ambient pressure as origin for the pressure scale . Phases with long range order are shaded. $T_\rho$ and $T^\star$ mark the onset of 1D charge localization and 1D/2D deconfinement respectively. The dashed line between 2D and 3D regimes defines the low temperature domain for 3D coherence. Right. Pressure dependence of the transport activation. The activation for the
$c$ axis transport ($\Delta_c$) although decreasing under pressure survives up to high pressures while the longitudinal transport ($\Delta_a$) is no
longer activated  above 
15 kbar when a dimensional cross-over occurs at the  temperature $T^\star$ \cite{Auban04}. }
\label{generic2.pdf} 
\end{figure*}

In the mid-1980's, the isostructural  family comprising the sulfur molecule with the same series of monoanions  was investigated under pressure and it was realized that \tfx and  \tms2x salts both belong to a unique class  of materials forming the generic \tm2x phase diagram\cite{Jerome91}, fig(\ref{generic2.pdf}).    In particular,   \tmps , although  the most insulating compound of the phase diagram can be made  superconducting at low temperatures under a pressure  of 45 kbar\cite{Wilhelm01,Adachi00}. The discovery of \tmp6 \textit{via} the study of the generic phase diagram has thus contributed to the knowledge of 1D physics much more than the phenomenon of superconductivity in an organic conductor.
 Because of their particular crystal structures, such materials can be considered   at first glance  as protoptypes for 1D physics\cite{Bourbonnais98}, see also J. Friedel\cite{Friedel00} for a short historical summary on the role of correlations in organic conductors. 
 
 What 1D physics means in a nutshell, is the replacement of  Landau-Fermi quasiparticles  low lying excitations by decoupled spin and charge collective modes \cite{Schulz90,Voit95}.
 The model for 1D conductors  starting from a linearized energy spectrum for
excitations close to the Fermi level and adding the relevant Coulomb repulsions which are
responsible for electron scattering with momentum transfer 0 and $2k_F$ is known as the  popular
Tomonaga-Luttinger (TL)  model. In this model   the spatial variation of all correlation functions (spin  susceptibility at $2k_F$ or $4k_F$, $CDW$, Superconductivity) exhibit a power law decay at large distance, characterized by a non-universal exponent $K_{\rho}$ (which is a function of the microscopic coupling constants)\cite{Solyom79}. However an important peculiarity of \tm2x materials makes them different from the usual picture of TL conductors. Unlike two-stacks \tq materials, $\mathrm{(TM)_{2}X}$ conductors exhibit a band filing which is commensurate with the underlying 1D lattice due to  the 2:1 stoichiometry  imposing half a carrier (hole) per TM molecule. Consequently, given  a uniform spacing of the  molecules along the stacking axis the unit cell contains  1/2 carrier, \textit{i.e.} the conduction band is quarter-filled.  
However,
non-uniformity of the intermolecular spacing  had been noticed from the early structural studies of \tfx crystals\cite{Ducasse86} due to the periodicity of the anion packing twice the periodicity of the molecluar packing. This non-uniformity provides a dimerization of the intrastack overlap  which is important  in the sulfur series (prominent 1/2 Umklapp scattering)  but  still present although less developed in some members of the  
$\mathrm{(TMTSF)_{2}X}$ series. 

An important  consequence of the commensurate situation for $\mathrm{(TM)_{2}X}$ materials is the existence of localization channels due to electron-electron Umklapp scatterings with momentum transfer $4k_F$ (two particles scattering) or $8k_F$ (four particles scattering) corresponding to half or quarter-filled bands respectively\cite{Emery82,Giamarchi97}. This localization is a typical outcome of 1D physics in the presence of repulsive interactions and leads to  a charge gap $\Delta_{\rho}$ in the quasiparticle spectrum although no ordering is expected at any finite temperature for a purely 1D system.  The spin sector remains gap-less on account of the separation between spin and charge degrees of freedom in  1D conductors. Fortunately, most features of the 1D localization and the physical properties of this electron gas can be studied by transport and optical properties of the materials. The most salient effect is  the existence of  a minimum of  resistance at $T_{\rho}$ below which the resistance becomes activated. According to the theory of the commensurate 1D electron gas  worked out by a number of authors, using the extended Hubbard, DMRG and bozonisation models, ($\mathit{see}$ references\cite{Giamarchi04,Giamarchi08} for further reading) it has been shown that both Umklapp mechanisms compete for the establishment of the charge gap, 1/4  and 1/2 Umpklapp scatterings leading to charge ordered and Mott insulators respectively\cite{Tsuchiizu01}.

As long as the carriers are localized on individual stacks, the Mott gap prevents the formation of electron hole pairs and in turn  any hopping to neighboring stacks\cite{Bourbonnais91,Giamarchi04}. Therefore, even if a finite interchain coupling $t_b$ already exists according to the band structure calculation in real materials, it is not recognized as a pertinent quantity. The carriers are confined on individual stacks as long as $\Delta_{\rho}$ remains finite. The 1D confinement is affected by pressure due first to the weakening of the correlations and second by the interplay between the 1D localization and the single particle interchain hopping $t_b$. Once the interchain single particle hopping (increased by pressure) reaches the order of magnitude of the Mott gap, confinement ceases and the existence of a quasi-1D Fermi surface becomes meaningful\cite{Giamarchi97,Bourbonnais99,Biermann01}. In some respects the deconfinement observed under pressure around 15 kbar on fig(\ref{generic2.pdf}) is the signature for a crossover from strong to weak coupling in the generic phase diagram. The N\'eel phase turns into  an ordered  weak antiferromagnet (spin density wave) at  the deconfinement pressure.

On the left side of the generic diagram the Mott localization, the consequence of Umklapp scattering as a one dimensional effect in a  TL electron gas  does not establish long order. Various types of range orders occur only at low temperature in this 1D Mott-localized sector.

On  the very left side of the generic diagram on fig(\ref{generic2.pdf}) a phase transition towards a long range ordered insulating phase has been observed   in the charge-localized temperature sector. This phase at low temperature has been  ascribed, according to NMR data, to the onset of a charge  disproportionation between molecules on the molecular chains\cite{Chow00a}. Since the charge of this low temperature phase (\textit{see}, the CO state on the left side of the generic \tmps  diagram, figure (\ref{generic2.pdf}), is no longer uniform ferroelectricity can be expected as shown by a signature in dielectric measurements\cite{Nad00}. The stability of the CO state (often called a Wigner state) is a direct consequence of the long range nature of the Coulomb repulsion which, in terms of the extended Hubbard model, amounts to  finite on-site U and second-neighbours V repulsions. Increasing pressure, spins localized on dimers of molecules couple to the lattice and formed a spin-Peierls ground state evolving possibly through a quantum critical point\cite{Chow98}  into a N\'eel antiferromagnetic phase under pressure.

\subsubsection{The generic diagram studied by transport}
Regarding the upper part  of the diagram on  fig(\ref{generic2.pdf})  a metal-like behaviour of the  longitudinal resistance is still be oberved at $T$ larger than  $\Delta_{\rho}$ leading to a  resistance displaying a power law $\rho (T) \approx T^{\theta}$  ($\theta >0$). Experimental studies  have revealed such a metallic behaviour for the  resistance either in \tfx under pressure\cite{Auban04,Degiorgi06} or  in \tsx even at ambient pressure\cite{Jerome82}. What has been found is a sublinear exponent, namely $\alpha = 0.93$ for the constant volume temperature dependence  of  the \tmp6 resistance\cite{Auban04,Degiorgi06}. 

Transport  experiments can  be explained within the theoretical framework of commensurate conductors with  commensurability  one or two. However, in the case of half-filling  (commensurability one) the Luttinger parameter $K_{\rho}$ should be very close to unity (very weak repulsive interactions) since $\theta = 4K_{\rho}-3$. A situation which implies weak coupling $K_{\rho}\approx 1$ (nearly free electrons)  would be hard to reconcile with the exchange enhancement of the spin susceptibility, an other sign for the role of correlations\cite{Bourbonnais99}. On the other hand, when quarter-filled Umklapp scatterings prevail at high temperature $\theta = 16K_{\rho}-3$ and consequently  $K_{\rho} = 0.23$ according to the  data of \tmp6. 
 
Other arguments in favour of a predominance of the quarter-filled Umklapp scattering above $T_{\rho}$ are given by two additional experimental features. 

First, the  phase diagram of an organic salt \edt \\  in which for symmetry considerations (the absence of inversion center between adjacent molecules along a stack and the existence of a glide plane symmetry\cite{Heuze03}) dimerization is prevented, hence 1/4- Umklapp scattering  is the only channel left to explain the carrier localization in this 1D conductor. In spite of the different structures  the  high temperature limit of  phase diagrams of \edt\,and \TM\, are fairly similar\cite{Auban09}.

Second,  the temperature dependence of the transverse conduction along  the least conducting  $c$ direction  which   displays an insulating behaviour below room temperature going through   a maximum around $T^{\star} \approx 80-100$ K and becoming metallic at lower temperatures. 
In this metallic regime the transverse resistance is remaining several orders of magnitude above the  Mott-Ioffe critical value which is considered as the limit
between metal and insulating-like transport \cite{Mott74}. 

As long as coherence is not established between 2 D planes, transverse transport requires  tunneling of Fermions ({\it at variance}\, with the longitudinal transport which is related to  1D collective modes)  between neighbouring chains and therefore probes the amount of quasi particles (QP) existing close to Fermi level.
The insulating character of the  transverse transport at high temperature can thus be interpreted as the signature of a non
Fermi-Landau behavior \cite{Moser98}. When  transport along the
$c$-direction is incoherent, \textbf{transverse conductivity probes the physics of the $a-b$ planes} namely, the physics of the weakly interacting Luttinger chains in the $a-b$ planes. The resistivity along the least conducting direction depends on the one-electron spectral function of a single chain  and reads $\rho_{c}(T) \approx T^{1-2\alpha}$\cite{Georges00} where $\alpha$ is related to\, $K_{\rho}$,  ($\alpha =
\frac{1}{4}(K_{\rho} + 1/K_{\rho}-2)$. It is again a $K_{\rho}$ of about 0.25-0.30 which is found experimentally in \tmp6\cite{Moser98}.

\subsubsection{1D-2D crossover}
The temperature $T^{\star}$ where the $c^{\star}$-axis transport switches from an insulating
to a metallic temperature dependence corresponds to a cross-over between two regimes, see fig(\ref{generic2.pdf}); a high temperature regime
with no QP weight at  Fermi energy (possibly a TL liquid in the 1D case) and another regime in which the QP
weight increases with decreasing temperature. This picture does not necessarily imply that the transport along
the $c$-direction must also become coherent below the cross-over since the $c$-axis transport may  remain
incoherent with a progressive establishment of a Fermi liquid   in  $a-b$ planes at temperatures below
$T^\star$.   
Consequently, the temperature dependence of transport properties along  \textit{a} and \textit{c}-axes in the 1D regime above
$T^\star$  lead to a  determination of
$K_{\rho}$ of order 0.23  in \tmp6 and \tmps under a pressure of 12-15 kbar respectively. The Luttinger parameter is much smaller in  \tmps at ambient pressure\cite{Auban04}, ($K_{\rho}$=0.18) due to the enhancement of correlations in the left region of the phase diagram, fig(\ref{generic2.pdf}). Optical data measured throughout  the entire generic diagram reach fairly similar conclusions\cite{Pashkin06}.

The decrease of the charge gap under pressure, fig (\ref{generic2.pdf}), is due to a weakening of intra chain correlations moving from left to right in the diagram. Carriers are confined on the 1D chains and the  transverse coupling renormalized by intrachain interactions is not pertinent\cite{Bourbonnais84}. 

 However, when this  gap becomes of the order of the bare kinetic transverse coupling a close interplay between this gap and the transverse coupling occurs giving rise to the sharp suppression of the localization observed around 15 kbar in \tmps . At  higher pressures the transverse coupling becomes pertinent and approaches the bare coupling which is  increasing only  smoothly under pressure. The deconfinement of the carriers is observed around $T^\star$. Below this deconfinement temperature    charge excitations lose  their 1D character and resemble more and more  what is expected in Fermi liquids (quasiparticles), leading in turn to a quadratic temperature dependence for the longitudinal resistivity.
However,
 electron excitations of this ''Fermi liquid'' retain  a low energy gap in the far infra-red spectrum in which  most of the
oscillator strength is carried by states above the gap coexisting  with a very narrow and intense zero frequency peak in the conductivity\cite{Cao96,Schwartz98}. 

\subsubsection{2D-3D crossover}

 A three dimensional anisotropic coherent picture seems to prevail at low temperature in \tms2x compounds  according
to  the Kohler's rule \cite{Cooper86} and  angular magnetoresistance oscillations observed in \tmc and \tmp6 under pressure\cite{Kang92,Osada91,Danner94,Sugawara06}. In addition, optical reflectance data of light polarised along $c^{\star}$ support the existence for \tmc of a weak Drude behaviour in the liquid helium temperature domain when $k_{B}T<t_{\perp c}$ \cite{Henderson99}.
However, the upper limit for the temperature domain of 3D coherence has been established comparing the temperature dependence of the resistivity along $a$ and  $c^{\star}$. This  3D to 2D crossover domain is defined by the temperature above which  temperature dependences of both components of transport cease to be identical\cite{Auban11a}. 
The 3D coherence  regime is plotted on the generic diagram of  figure(\ref{generic2.pdf}). 

\subsubsection{The generic diagram studied by optics}
\begin{figure*}[htbp]	
 \centerline{\includegraphics[width=1\hsize]{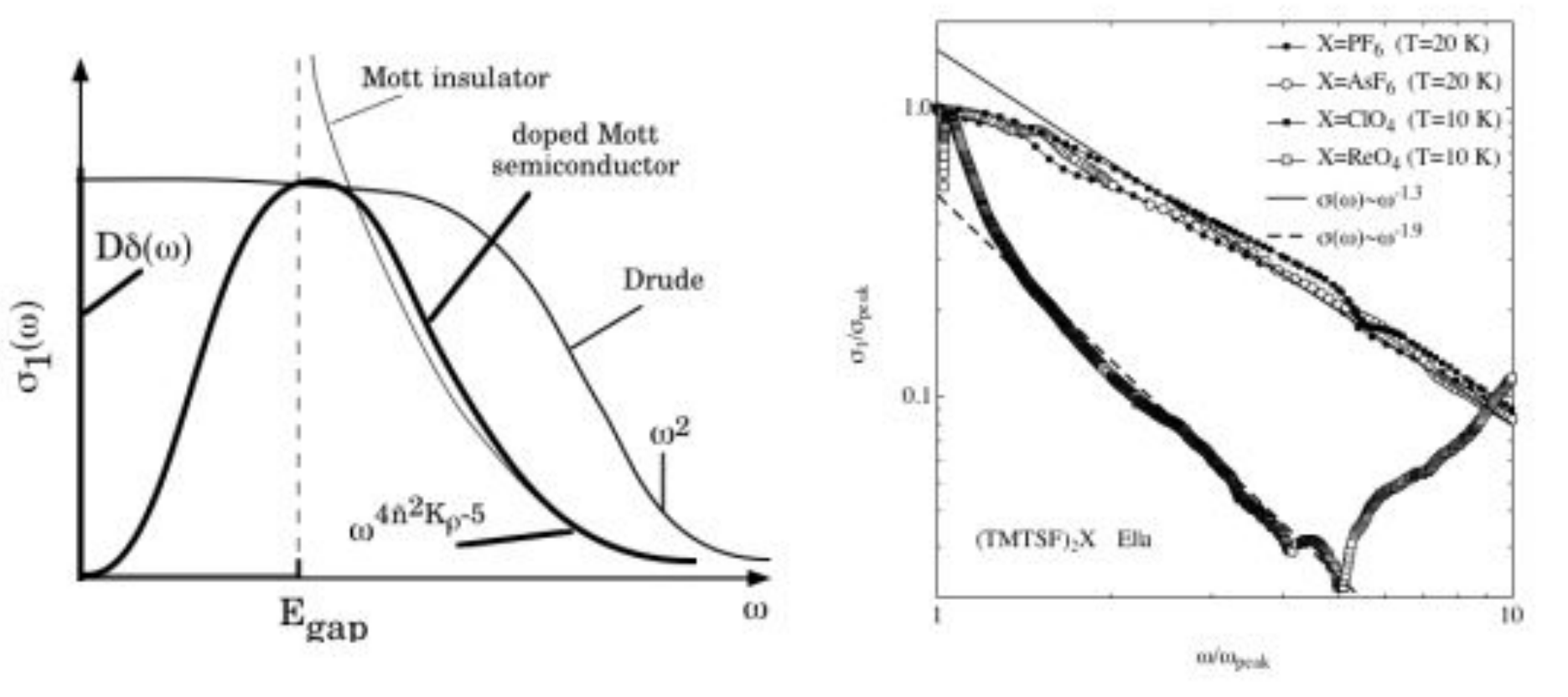}}
\caption{(Left ) Optical conductivity of a Mott insulator and of a doped Mott semiconductor\cite{Giamarchi08}. The power law decay of the optical conductivity at high frequency depends on the Luttinger parameter $K_{\rho}$. (Right) Power law behaviour of the optical conductivity for various \TM\, compounds measured in the metallic phase at low temperature (except for \R which is an insulator due to the anion ordering below 180 K), according to references\cite{Vescoli98,Zwick00}. }
\label{opticsTM2X.pdf} 
\end{figure*}

It is remarkable that most features of the generic phase diagram presented on fig(\ref{generic2.pdf}) are very well recovered by the study of low frequency excitations  via the far infra red optical conductivity investigations performed on either different materials of the \TM series or on a given compound  under pressure. 
Compounds on the left of the diagram exhibit a charge gap $E_{gap}$ due to the 1D Mott localization, $E_{gap}$=1100  and 375 K for \tmps and \tmttfbr respectively\cite{Degiorgi06}.
These values for the optical gap are indeed in fairly good agreement with the DC transport activation energy shown on fig(\ref{generic2.pdf}) since $2\Delta_{activation}=E_{gap}$. Moving towards right in the phase diagram, the optical evolves and resembles at low temperature the spectrum expected in doped Mott semiconductor, see fig(\ref{opticsTM2X.pdf}), namely, a high frequency component at $\omega > t_b$ carrying most of the oscillator strenght, a correlation pseudogap located around 275 K and a narrow Drude peak carrying about 1\% of the total oscillator strenght\cite{Vescoli98}.
The high temperature 1D TL properties of the system are consistently observed through the power law frequency dependence of the conductivity at high frequency above the correlation pseudogap, $\sigma(\omega)\approx \omega^{- {\gamma}}$ with $\gamma= 1.3$ \cite{Schwartz98}. Such an exponent is actually in good agreement with transport data of the selenide series since $K_{\rho}$ = 0.23 is recovered given the known relation $\gamma$=5-16$K_{\rho}$ when quarter-filled band Umklapp scattering is dominant\cite{Giamarchi97}.

The investigation of the pressure and temperature dependence of the optical conductivity of \tmps and \tmp6\cite{Pashkin06,Pashkin10} has also identified for \tmps a Mott deconfinement when the Mott gap is approximately twice $t_b$ and a dimensional crossover at $T^{\star}$ for both systems  revealed by  the development of a Drude type response  along $b$.

 Although not much work has been devoted to the optical response along the $c^{\star}$ direction the study performed on \tmc by Henderson \textit{et-al}\cite{Henderson99} indicates that this compound is a 2D metal above 10K and that a small Drude response emerges at lower temperature characterisitic of a 3D anisotropic electron gas.

\section{The last five years}
\label{sec:3}
At the turn of the 21$^{st}$ century, the prominent role of AF fluctuations in the metallic phase next to  the $SDW$ phase and the proximity between superconductivity and the magnetically ordered phase were both  well established experimentally\cite{Creuzet87,Bourbonnais08}. In addition, one was wondering about a possible connection between magnetism and electron pairing but even the symmetry of the $SC$ order parameter was a very controversial topic.

On the one hand early NMR studies had concluded to nodal superconductivity in \tmc \cite{Takigawa86} \textit{at variance} the with more recent Knight shift measurements in \tmp6  suggesting triplet \textit{p}-type pairing\cite{Lee05}. This latter possibility was actually in line with the claim of superconductivity surviving for certain orientations of the magnetic  field beyond the  field set by the Pauli limiting field $\mu_{0}H_{P}\sim 2.5$ T  where ordinary singlet pairs would become unstable on account of the Zeeman splitting for a \tc of 1.2 K. On the other hand a thermal conductivity study of \tmc concluded to a nodeless superconductivity\cite{Belin97} not necessarily associatd to \textit{s}-wave superconductivity as odd parity such as a \textit{p}-wave gap could explain the behaviour of temperature dependence of thermal conductivity. 

The clue could not come from a single experiment but from a coordinated experimental program with NMR in the superconduting state at UCLA\cite{Shinagawa07}, a study of the role of non magnetic impurities on \tc at Orsay\cite{Joo05}, field angle resolved calorimetry at Kyoto\cite{Yonezawa12} and the study of transport in the metallic phase surrounding superconductivity at Sherbrooke\cite{Doiron09,Doiron10}. These questions will be addressed in the rest of this paper.
\subsection{NMR in the superconducting phase}
 Regarding the spin part of the $SC$ wave function, a triplet pairing was first claimed from a divergence of the critical field \hc2 exceeding the Pauli limiting value reported at low temperature in \tmp6 under pressure when $H$ is applied along the $b'$ or $a$ axes \cite{Lee97} and from the absence of any change in the $^{77}\mathrm{Se}$ Knight shift at \tc \cite{Lee01}. 

\begin{figure*}[htbp]	
 \centerline{\includegraphics[width=0.7\hsize]{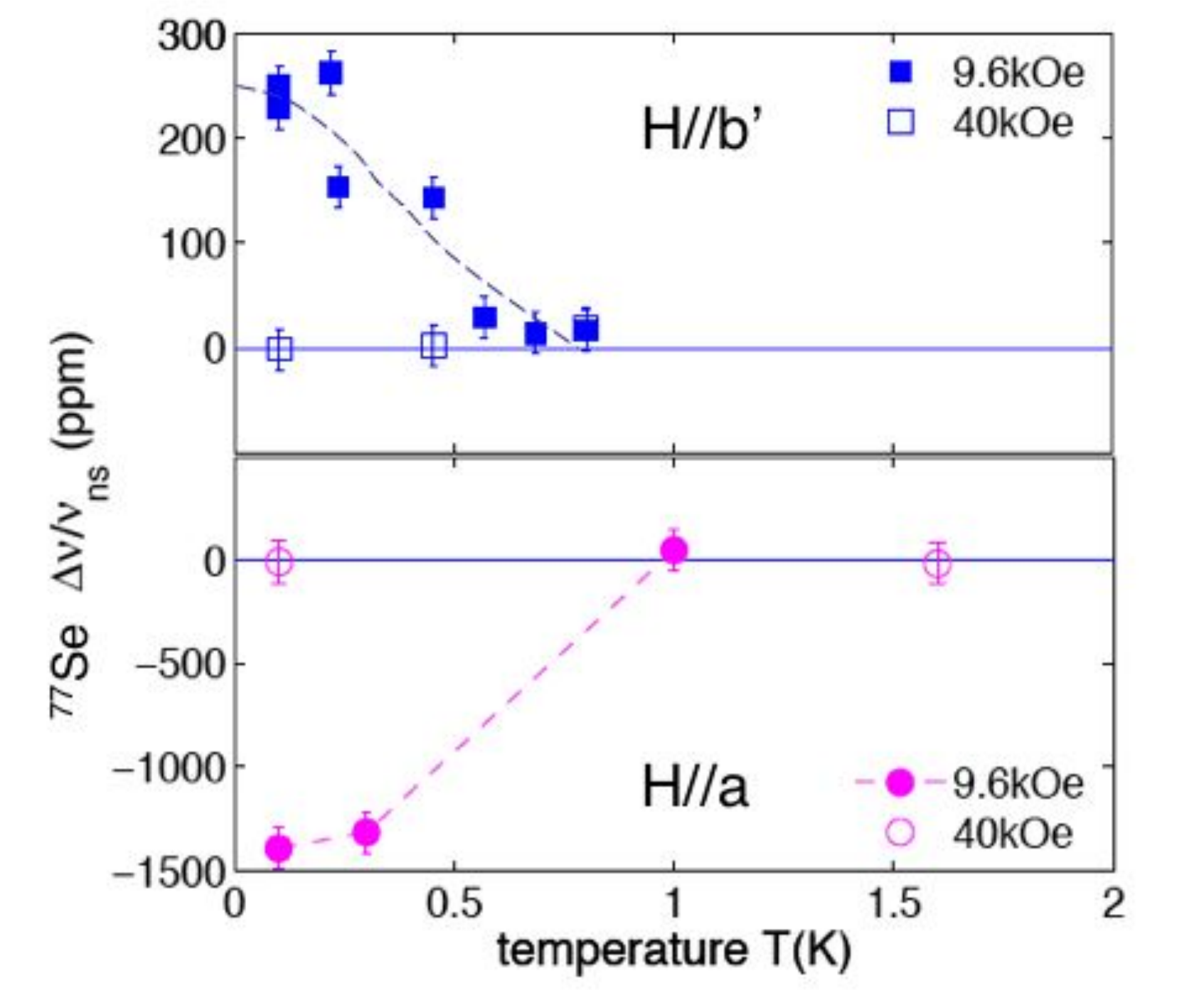}}
\caption{ $^{77}Se$ Knight shift vs $T$ for \tmc, for $H// b'$ and $a$, according to reference \cite{Shinagawa07}. The sign of the variation  of  Knight shift at \tc depends on the sign of the hyperfine field. A variation   is observed only under low magnetic field. The difference of \tc for the two field orientations is due to the anitropy of \hc2.  The field of $ 4T$ is still lower than the critical field derived from the onset of the resistive transition (\hc2$\approx 5T)$, \textit{vide} Sec().  }
\label{KnightshiftClO4.pdf} 
\end{figure*}

However more recent experiments performed at fields lower than those used in reference\cite{Lee01} for the work on \tmp6 did reveal a  clear drop of the  $^{77}\mathrm{Se}$ Knight shift tbelow \tc in the compound \tmc \cite{Shinagawa07}. These new data  provided a conclusive evidence in favour of singlet pairing, (Fig.\ref{KnightshiftClO4.pdf}). In addition, a step increase of the spin lattice relaxation rate versus magnetic field  for both field orientations $\parallel a$ and $b'$ provided  evidence for a sharp cross-over or even a phase transition  occuring at low temperature under magnetic field  between the low field \textit{d}-wave singlet phase and a high field regime exceeding the paramagnetic limit $H_p$ being either a triplet-paired state\cite{Shimahara00,Belmechri07} or an inhomogenous Fulde-Ferrell-Larkin-Ovchinnikov state\cite{Fulde64,Larkin65}.

\subsection{Organic superconductivity and non magnetic impurities}
The recent finding of a \tc strongly affected by the value of the elastic electron lifetime in the presence of non magnetic lattice defects has conclusively ruled out the \textit{s}-wave hypothesis and suggested the existence of nodes with zeros of  the $SC$ gap on the Fermi surface\cite{Joo05}.

Leaving the cation
stack uniform in order to preserve the  electronic structure, a soft way of introducing non magnetic disorder scattering centers can be achieved by playing  on the anion stacks. The role of the anion
stack on the ground state is enhanced as soon as the anion  located at an inversion center of the
structure  does not possess a central symmetry\cite{Pouget87}. This is the case in particular for tetrahedral anions
such as $\mathrm{X=ClO_4}$ which order  at low temperature ($T_{AO}$=24K) in line with 
entropy minimization. As anion reorientation requires a  tunneling process between two states at equal energy but
separated by  a large potential barrier, the dynamics of orientation is a slow process at low
temperature and becomes even slower under pressure when the tunneling barrier incraeses. Hence, for  samples slowly cooled through $T_{AO}$ (in the so-called  R-state) the orientation of the anions is
uniform along the stacking axis  but alternate along the  b-direction leading in-turn to a doubled
periodicity with a concomitant opening of an energy gap $\Delta_{anion}$ on the Fermi surface at  $\pm \pi /2b$ and the creation of two
sheets of open Fermi surfaces at +$k_F$ and -$k_F$ respectively. Fast cooled samples on the other hand reach low
temperature in a state (the Q-state) where orientational  disorder is frozen-in. The superconducting state arises at a depressed \tc in a sample where homogenous anion-oriented domains coexist with anion disorder outside of these domains\cite{Tomic83b}.  When the cooling rate becomes fast enough in a
Q state full disorder of the anions  is preserved ($\xi_{anion}<v_{F}/\Delta_{anion}$) and the single-sheet warped Fermi surface of the high temperature
sructure prevails at low temperature leading in turn to the instability of the metallic phase against a $SDW$ ground
state at $T_{SDW}$=5K\cite{Tomic83b}. Furthermore, it has been shown that neither the Pauli susceptibility\cite{Tomic83a} nor the density of
states\cite{Garoche82} of the normal phase are affected by the orientational disorder introduced by the fast cooling procedure as long as a superconducting
ground state is observed.  

An other  approach for the introduction of anionic disorder has been successful with the synthesis of an anionic solid solution
involving anions of similar symmetry. 

\begin{figure*}[htbp]	
\centerline{ \includegraphics[width=0.7\hsize]{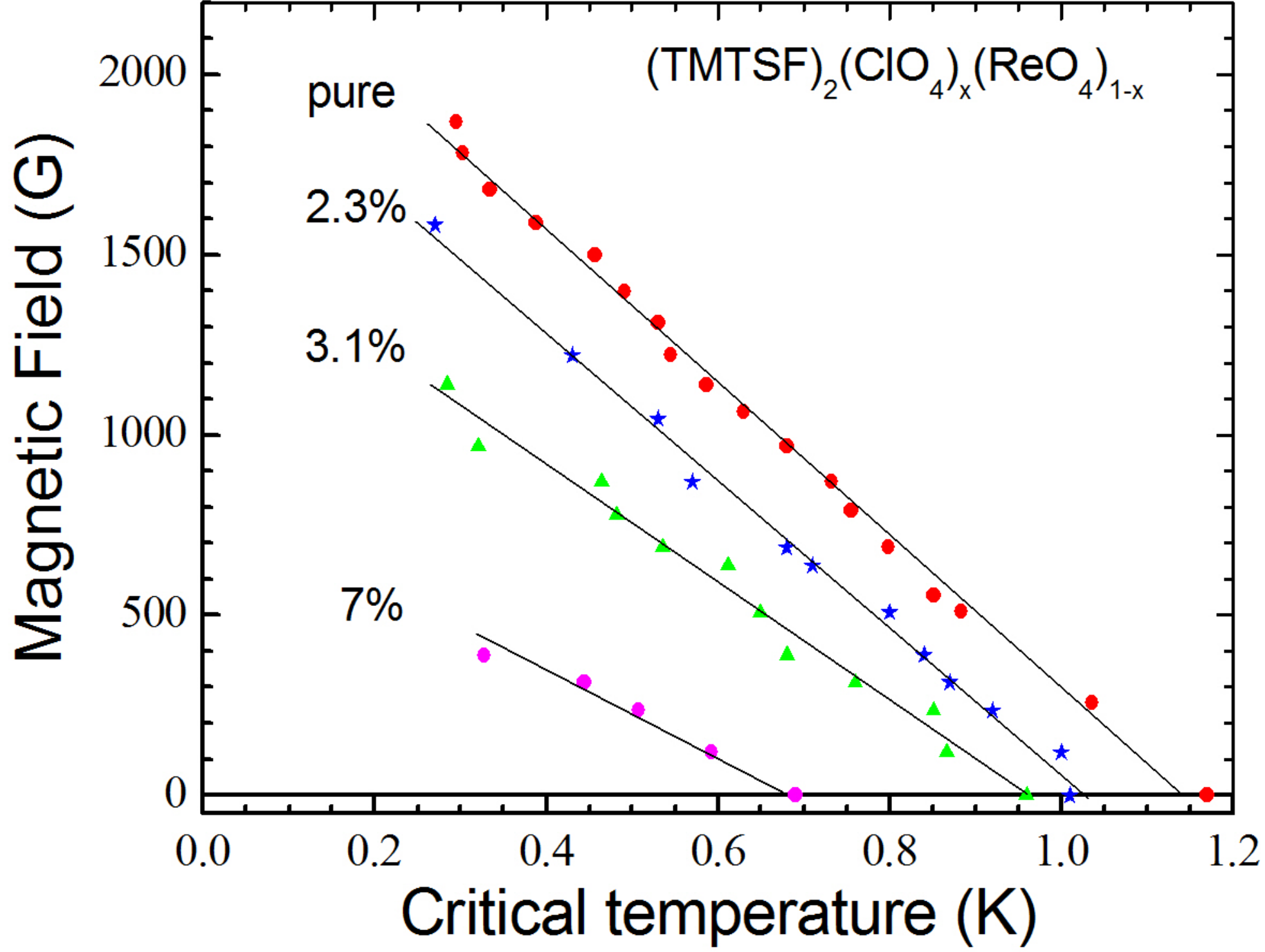}}
\caption{Critical field \hc2 for \tmx alloys with different concentrations of \R all slowly cooled in the R state. }
\label{HcTc.pdf} 
\end{figure*}

 The early studies by
Tomi\'c \textit{et-al}
\cite{Tomic83b} in \tmx \, have shown that both the low temperature conductivity and the  transition towards
superconductivity are very strongly affected by alloying although X-ray investigations have revealed
that long range order is preserved up to 3\% $\mathrm{ReO_{4}^{-}}$ with a correlation length $\xi$$_a$  $>$ 200\AA
\cite{Ravy86}.
Single crystals of  this solid solution  with $x$ in the range $0\leq x\leq 0.17$ have been
prepared with the usual electrocrystallization technique and studied for their behaviour at low temperature in a dilution fridge. Up to about 6\% $\mathrm{ReO_{4}^{-}}$ nominal concentration a $SC$ transition is observed although depleted from the \tc in pristine samples. Meanwhile, an insulating ground state is stabilized above 10\% $\mathrm{ReO_{4}^{-}}$ or so.
The criticial temperature has been determined from the temperature where \hc2 reaches a zero value, see figure (\ref {HcTc.pdf}).

The data of fig (\ref{HcTc.pdf}) show that   the slope $dH_{c2}/dT$ follows $T_c$ as this latter quantity is decreased upon alloying. This is the behaviour expected in a clean type II superconductor\cite{Gorkov85}, but in such a situation $T_c$ itself should not be affected by  non magnetic impurities. 

\begin{figure*}[htbp]	
\centerline{ \includegraphics[width=0.7\hsize]{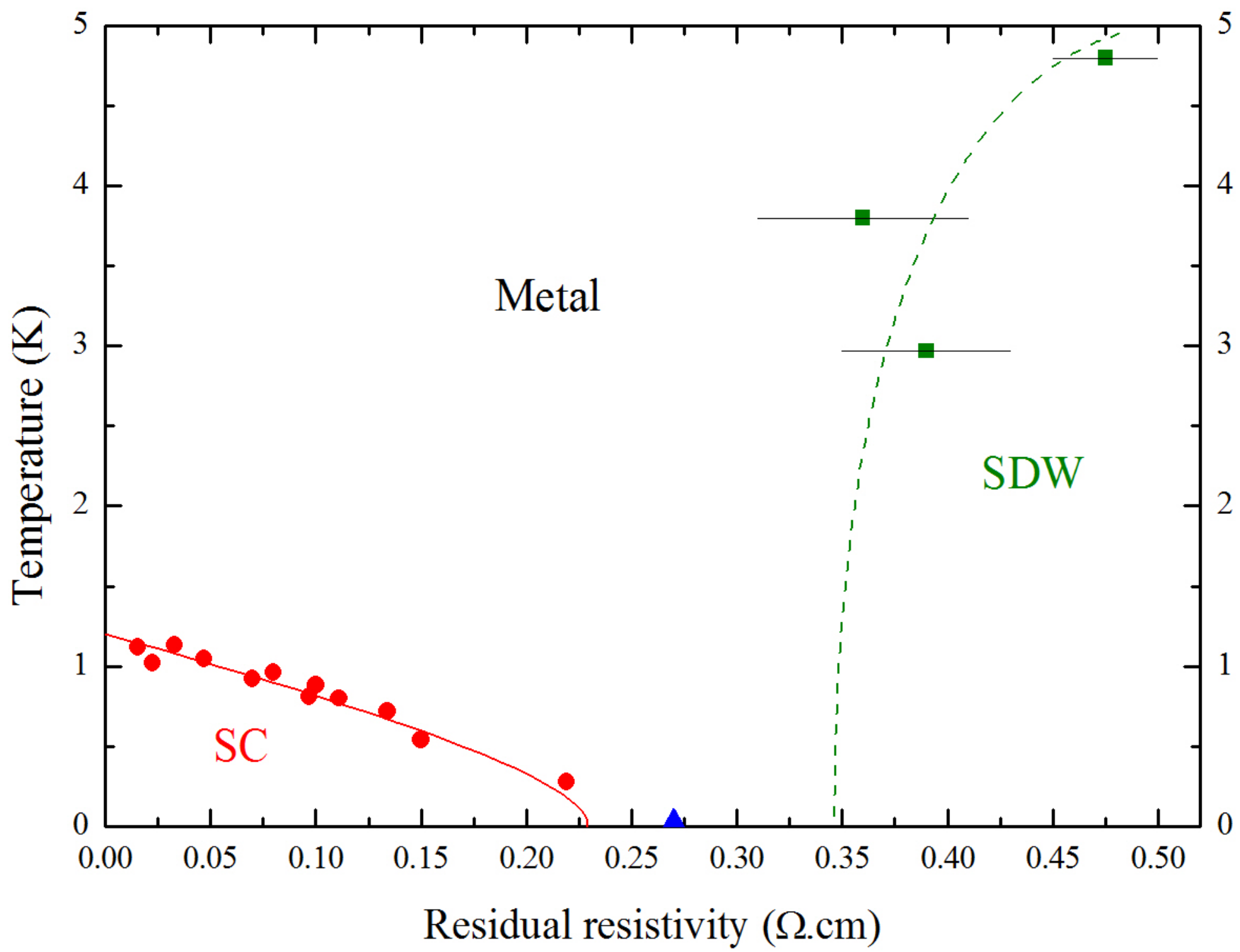}}
\caption{Phase diagram of \tmx \, governed by  non magnetic disorder acording to reference\cite{Joo05}. All  circles  refer to  very slowly cooled samples in the
R-state with different \R contents. The triangle at $T=0 K$ for the sample of $\rho_0=0.27(\Omega cm)^{-1}$ is metallic down to the lowest temperature of the experiment. When an upturn of the resistivity is observed signaling the onset of a $SDW$ state no sign of superconductivity can be noticed down to the lowest temperature. The continuous line (red) is a fit of the data with the Digamma function and $T_{c}{^0}=1.23K$}.
\label{TcRho0c.pdf} 
\end{figure*}

 At higher
impurity content, ($x$=15\%) a new behaviour has been observed. The anions
 still order below $T_{AO}$ remaining of the order of 24K as indicated by the concomitant drop  of the residual resistivity
(although smaller than what is observed in pristine samples). An upturn of
the resistivity noticed at low temperature in the R-state can be ascribed
to the onset of an insulating ground state at 2.35 K (as defined by the maximum
of the logarithmic derivative of the resistance \textit{versus} \textit{T}). A summary of this investigation is displayed on fig (\ref{TcRho0c.pdf}) where \tc is plotted against the residual (elastic) resistivity  $\rho_0$. 

\begin{figure*}[htbp]	
 \centerline{\includegraphics[width=0.8\hsize]{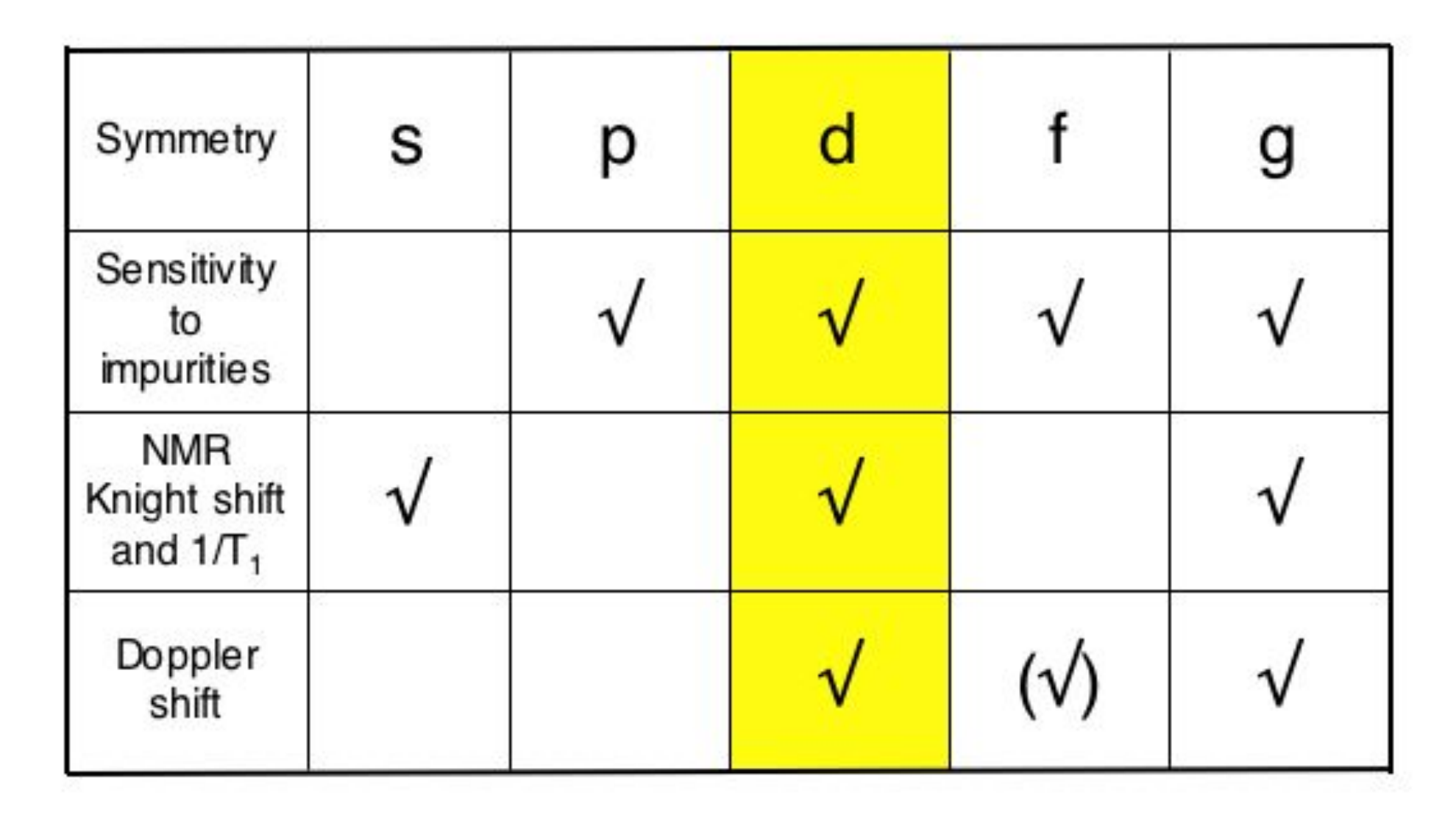}}
\caption{Possible gap symmetries agreeing with the different  experimental results. The \textit{d}-wave  (or \textit{g}-wave) symmetry  is the only symmetry agreeing with all experiments, (yellow column on line).}
\label{Nodes.pdf} 
\end{figure*}

However, the  determination of the elastic contribution is by no means an easy task. As the field of organic superconductors were developing over the years several situations regarding the behaviour of the resistivity of the metallic phase close to \tc could be obtained ranging from the classical $T^2$ Fermi liquid behaviour to a linear temperature dependence and even to a sublinear temperature dependence in the vicinity of \tc. The residual resistivity on fig (\ref{TcRho0c.pdf}) has been determined by the extrapolated value to zero temperature  of the linear behaviour observed below $4K$.
The suppression of $T_c$
is clearly related to the enhancement of the scattering rate in the solid solution. 
Since the additional scattering cannot be ascribed to magnetic scattering according to the  EPR checks
 showing no additional traces of localized spins in  the solid solution, the data in figure (\ref{TcRho0c.pdf}) 
 cannot be reconciled with the picture of a superconducting gap keeping a constant sign over the whole
$(\pm k_F)$ Fermi surface. They require a picture of pair breaking in a superconductor with an unconventional gap
symmetry. The conventional pair breaking theory for magnetic impurities in usual superconductors has been generalized to
the case of non-magnetic impurities in unconventional materials and the correction to $T_c$ obeys the following relation \cite{Maki04,Larkin65},
\begin{eqnarray}
\ln\Big(\frac{T_{c}^0}{T_{c}}\Big)=\psi\Big(\frac{1}{2}+\frac{\alpha T_{c}^0}{2\pi T_{c}}\Big)-\psi\Big(\frac{1}{2}\Big),
\end{eqnarray}
with $\psi(x)$ being the Digamma function, $\alpha = \hbar /2 \tau k_{B}T_{c}^0$ the depairing parameter, $\tau$
the elastic scattering time 
and $T_{c}^0$ the limit of $T_c$ in the absence of any
scattering. The experimental data follow the latter law with a good accuracy with 
$T_{c}{^0}=1.23K$. The influence of non magnetic impurities on the superconducting phase implies the existence of  positive as well as negative values for the $SC$  order parameter. It precludes the usual case of \textit{s}-symmetry but is unable to 
 discriminate between  two possible options namely,  singlet-\textit{d (g)} or triplet-\textit{p (f)}\cite{Nickel05}, see fig(\ref{Nodes.pdf}). 

\subsection{Doppler shift  and paramagnetic limitation}
The  response of superconductivity to non magnetic impurities and the loss of spin susceptibility at \tc should be sufficient to qualify the $d$-wave  alternative as the likely one among the various possiblities displayed on fig (\ref{Nodes.pdf}). In such a case, nodes of the gap should exist  on the Fermi surface and could be located.
\begin{figure*}[htbp]	
 \centerline{\includegraphics[width=1\hsize]{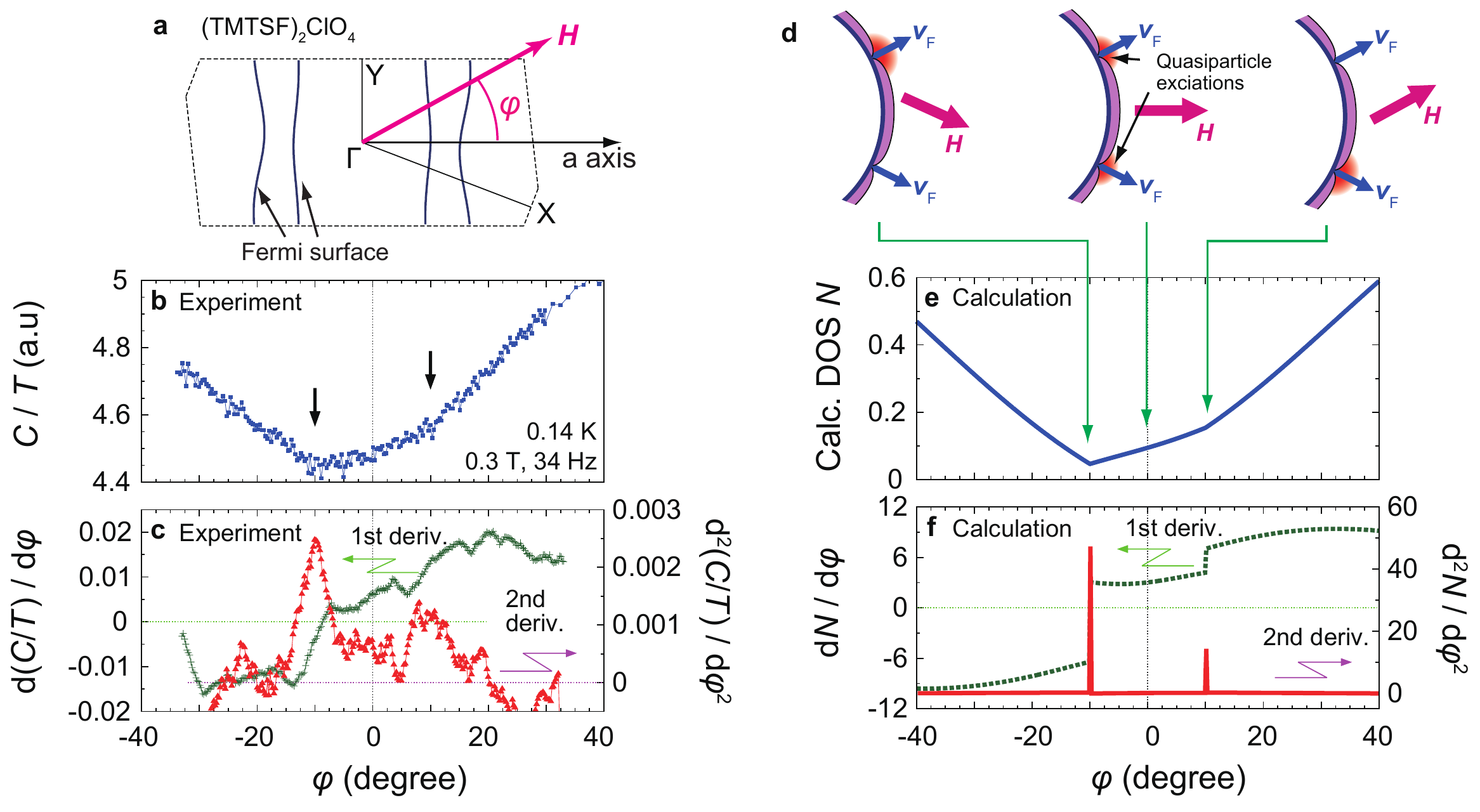}}
\caption{a-  Fermi suface of \tmc at low temperature in the presence of anion ordering\cite{Lepevelen01}. b- Angular dependence of $C(\phi)$ at 0.14 K and 0.3 T according to reference\cite{Yonezawa12}. c- First and second derivatives of the  data displayed on a. d- Sketch of the influence of a magnetic field on the quasi particle DOS. The contribution of Doppler-shifted quasi-particles  to the DOS is minimum when  the magnetic field is  parallel to $\textbf{v}_F$ at one of the nodes (left or right) . e and f- Simulations of the specific heat in the vicinity of the angle $\phi=0$ using the band structure on fig(\ref{volovik-effect_for_Review_ProfJerome_.pdf}a). Arrows correspond to the different orientations of the magnetic field as shown on fig(\ref{volovik-effect_for_Review_ProfJerome_.pdf}).
} 
\label{volovik-effect_for_Review_ProfJerome_.pdf} 
\end{figure*}
This has been made possible via a  measurement of the quasi particle density of states in the superconducting phase performed in an oriented magnetic field\cite{Yonezawa12}, see figs (\ref{volovik-effect_for_Review_ProfJerome_.pdf}b and c). This experiment is based on a   Volovik's important remark\cite{Volovik93} that for superconductors with line nodes, most of the density of states in their mixed state under magnetic field comes from the superfluid density far outside the vortex cores. The energy of QP's in the superfluid density rotating around the vortex core with the velocity $\textbf{v}_s(\bot H)$ is Doppler-shifted by the magnetic field and reads,
 \begin{equation}
\delta\omega\propto \textbf{v}_s\cdot \textbf{v}_F(\textbf{k})
\end{equation}
\label{2}
 where $ \textbf{v}_F(\textbf{k})$ is the Fermi velocity at the $\textbf{k}$ point on the Fermi surface.
 The Doppler shift can contribute to the DOS in the superconducting state as long as  it remains smaller than the superconducting energy gap, \textit{i.e} only for those \textbf{k} states   located in the vicinity of gap nodes on   the Fermi surface. Hence,  the contribution of the Doppler shift is minimum when the Fermi velocity at a node and the magnetic field are parallel according to eq (2), see fig(\ref{volovik-effect_for_Review_ProfJerome_.pdf}d) and   should contribute to a kink in the rotation pattern of $C(\phi)$.
The Doppler shift has already been used extensively in the case of 2D superconductors expected to reveal line nodes probing the thermal conduction or the specific heat while the magnetic field is rotated in the basal plane\cite{Yamashita11,Sakakibara07}. For such 2D conductors  $\textbf{v}_F(\textbf{k})$ is usually colinear with \textbf{k}. Therefore,  the  angular resolved specific heat (or thermal conductivity) enables to reveal the positions of the gap nodes according to the angles corresponding to  minima of  specific heat (thermal conductivity)\cite{Yamashita11}.

The situation is somewhat peculiar for Q1D Fermi surface. There, anomalies in the rotation pattern of $C_v$ are expected at angles $\phi_n$ between $H$ and $\textbf{v}_F(\textbf{\textbf{k$_n$}})$
where  \textbf{k$_n$} corresponds to a nodal position  on the Fermi surface.

A simple model has been used to understand the  data of angular resolved $C_v$ of \tmc\cite{Yonezawa12}. The rotation pattern has been modeled  by,
\begin{equation}
N(\phi)\propto \sqrt{H/H_{c2}(\phi)}\Sigma_{n}A_{n}\mid sin(\phi - \phi_{n})\mid
\end{equation}
 where the first factor reproduces the anisotropic character of the critical field in the basal plane while  the weighted summation  over angles accounts for the existence of nodes at angles $\phi_n$ between the magnetic field and the special  \textbf{k$_n$} points where the Fermi velocity is parrallel to $H$.
 The experimental observation of a rotation pattern at low field and low temperature which is non symmetrical with respect to the inversion of  $\phi$ and the  existence of kinks for  $C(\phi)$  at $\phi=\pm 10^o$ on  fig (\ref{volovik-effect_for_Review_ProfJerome_.pdf}b) have been taken as  the signature of line nodes\cite{Yonezawa12}. According to the band structure calculation the location of the nodes on the Fermi surface becomes $k_y\sim\pm0.25b^*$. As  the Pauli limitation is concerned, the superconducting phase diagram of \tmc established by thermodynamic experiments indicates a value for the thermodynamic \hc2 along $a$ of 2.5 T for $H\parallel a$ much smalller than the expected value of 7.7 T for the orbital limitation derived from the measurement of the temperature derivative of \hc2 close to \tc for a clean type II superconductor\cite{Gorkov85}. On the other hand, along the two transverse directions orbital limitation is at work\cite{Yonezawa12}. The domain of field above the paramagnetic limit up to \hc2 derived from resistivity data where the specific heat recovers its normal state value  requires further experimental investigations but it might be the signature of some $FFLO$ phase as suggested by A. Lebed recently \cite{Lebed10}.

 \subsection{A Non-Fermi liquid metallic state}
 
Let us now address   a question which had been raised 30 years ago namely the non conventional behaviour of electronic transport in the metallic state in the neighborhood of the superconducting state. It was noticed that the resitivity displays a linear temperature dependence in the low $T$ regime becoming even sublinear in the very vicinity of the $SC$ phase\cite{Jerome80}. Such a behaviour is clearly not following the expectation for electron electron scattering in a Fermi liquid where$\rho\propto T^2$  even in a 2D electron gas.

An early explanation for this non-canonical behaviour had been attributed to 
 the existence of  paraconductivity in a quasi-one dimensional conductor above $T_c$  \cite{Schulz81}. The calculation of  Azlamazov-Larkin diagrams based on a time dependent Ginzburg-Landau theory provided, (i) a 3D domain restricted to the very vicinity of $T_c$ where the transverse coherence extends over several interchain distances $d$, \textit{ i.e}, $\xi_{\perp}/d>1$ with $\rho_{\parallel}^{3D}$ $ \alpha$ $ T^{-1/2} (ln T/T_c)^{1/2}$ leading in turn to the classical $1/(T-T_c)^{1/2}$ divergence of the paraconductivity and, (ii) a 1D regime where $\xi_{\perp}/d<1$ with $\rho_{\parallel}^{1D}$ $ \alpha$ $ T^{1/2} (ln T/T_o)^{-3/2}$ where $T_o$ represents a renormalized mean field temperature, about one third  of the actual mean field temperature \cite{Jerome82}. 
 
Although the comparison between  experiments and the theory for paraconducting fluctuations in  reference\cite{Schulz81} revealed a reasonable agreement in the 3D regime, 
it  failed to reproduce the correct temperature dependence in the range  4-15K.

The reinvestigation of transport of organic superconductors was necessary after similar non Fermi liquid behaviours with $T$- linear   laws had been reported in a lot of new superconductors discovered after 1986. This had been achieved in two steps, first  clarifying  transport properties experimentally and theoretically   below say, 20 K and second  understanding the vicinity of the superconducting transition   ($\Delta T/T_c \approx 0.2 )$
with sublinear transport behaviour.

\subsubsection{NFL,  AF fluctuations, normal phase spin susceptibility, transport and superconductivity}
A recent  extensive study of the transport properties has been carried on  in  Bechgaard salts superconductors, \tmp6\cite{Jerome80} and \tmc\cite{Bechgaard81} as a function of  pressure\cite{Doiron09,Doiron10}. This  study has focused on the electronic transport at low temperature in the low $T$ limit. 
\begin{figure*}[h]	
\centerline{ \includegraphics[width=0.7\hsize]{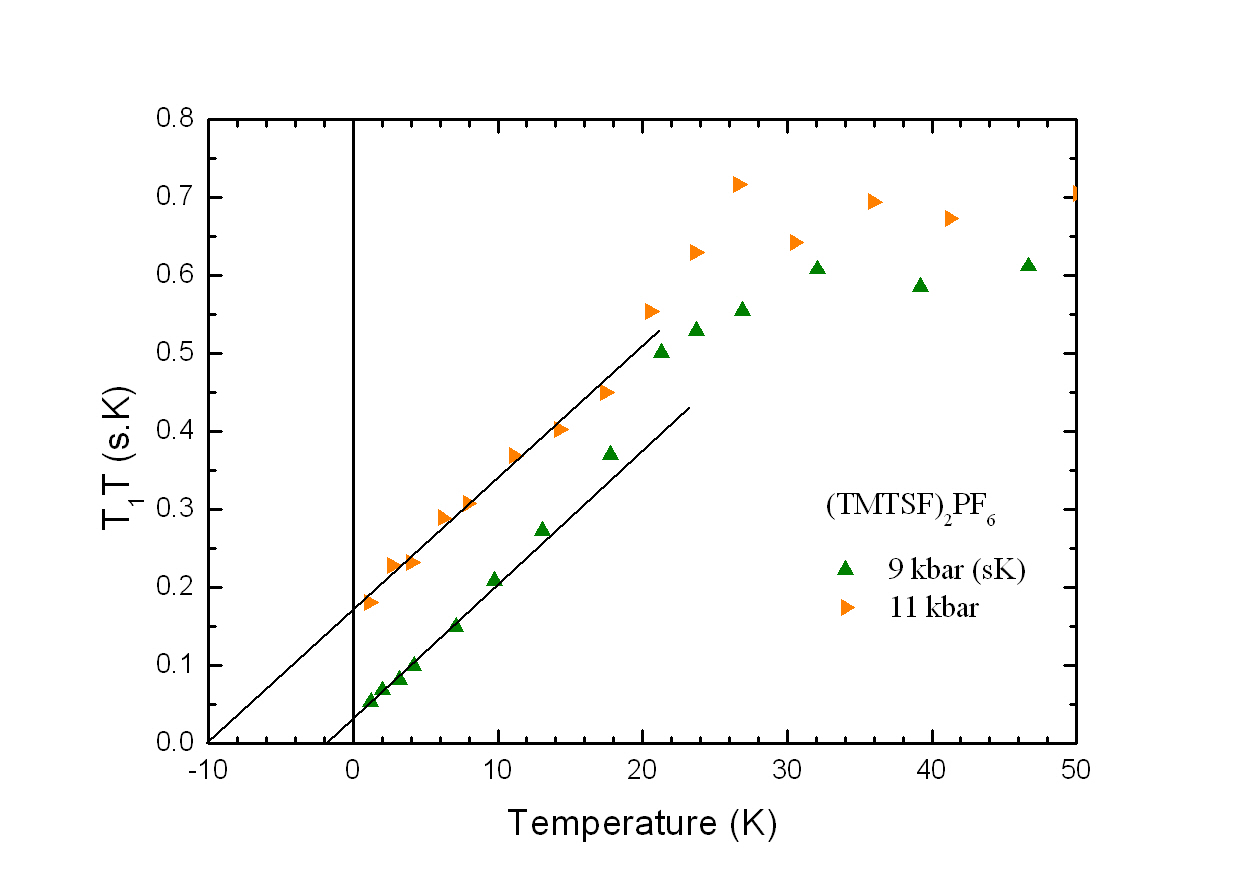}}
\caption{Plot of the nuclear relaxation versus temperature according to the data of reference \cite{Creuzet87}. The 2D AF regime is observed below $\approx 15 K$ and the small Curie-Weiss temperature of the 9 kbar run signals that this pressure is very close to a QCP probably hidden inside the $SDW$/$SC$ coexistence domain. The Korringa regime is recovered above $30 K$. }
\label{T1TvsT_PF6.jpg} 
\end{figure*}
Let us first notice that the magnetic metallic background of \tmp6 when superconductivity occurs under pressure although suppressed by the application of a magnetic field  applied along the $c^\star$ axis is far from being the expectation for a  Fermi liquid where a canonical Korringa law for the nuclear spin lattice relaxation namely $T_{1}T$ = constant is expected. Instead, for \tmp6 under 9 and 11 kbar  the law  $T_{1}T=C(T+\Theta)$ is followed as shown  on fig(\ref{T1TvsT_PF6.jpg}). The Korringa law is observed at high temperature say, above 25 K or so but the low temperature behaviour deviates strongly from the standard relaxation in paramagnetic metals.  

This Curie-Weiss behaviour of the relaxation is to be observed in a 2D fluctuating antiferromagnet\cite{Brown08,Wu05,Moriya00,Bourbonnais09} with the positive Curie-Weiss temperature $\Theta$ providing the energy scale of the fluctuations. It is vanishingly small when pressure approaches \pc but rises rapidly under pressure. The vanishing of this energy scale at \pc is thus a strong hint in favor of the existence of a magnetic quantum critical point (QCP) probably below   \pc but hidden by the $SDW/SC$ first order coexistence. On the other hand when $\Theta$ becomes large compared to $T$, the standard relaxation mechanism is recovered at low temperature in agreement with the very high pressure results. Consequently, a non standard behaviour is expected for the electron scattering at low temperature  in case it is governed by spin fluctuations.

Next, when transport of the metallic phase is examined a  log-log plot of the resistivity minus the residual resistivity at low temperature versus $T$ performed at every pressure 
 reveals at once a temperature dependence evolving from linear at low temperature to quadratic in the high temperature regime with a general tendency to become quadratic even at low temperature when pressure is well above the critical pressure \pc\cite{Doiron09}. It is clear that in the low temperature domain where linear and quadratic behaviours of the resistivity prevail the electron scattering cannot be due to phonons since the upper limit for electron-electron scattering should fall in the range below $T<\Theta_{Debye}^2/t_{\perp}$ \textit{e.g}, a temperature of the order of 200 K in a conductor such as  \tmp6.
 
 The existence of a linear temperature dependence of the resistivity is \textit{at variance} with the sole $T^2$ dependence expected from the electron-electron scattering in a conventional Fermi liquid. This is clearly  seen on a log-log plot of the resistivity versus $T$. 
\begin{figure*}[h]	
\centerline{ \includegraphics[width=0.5\hsize]{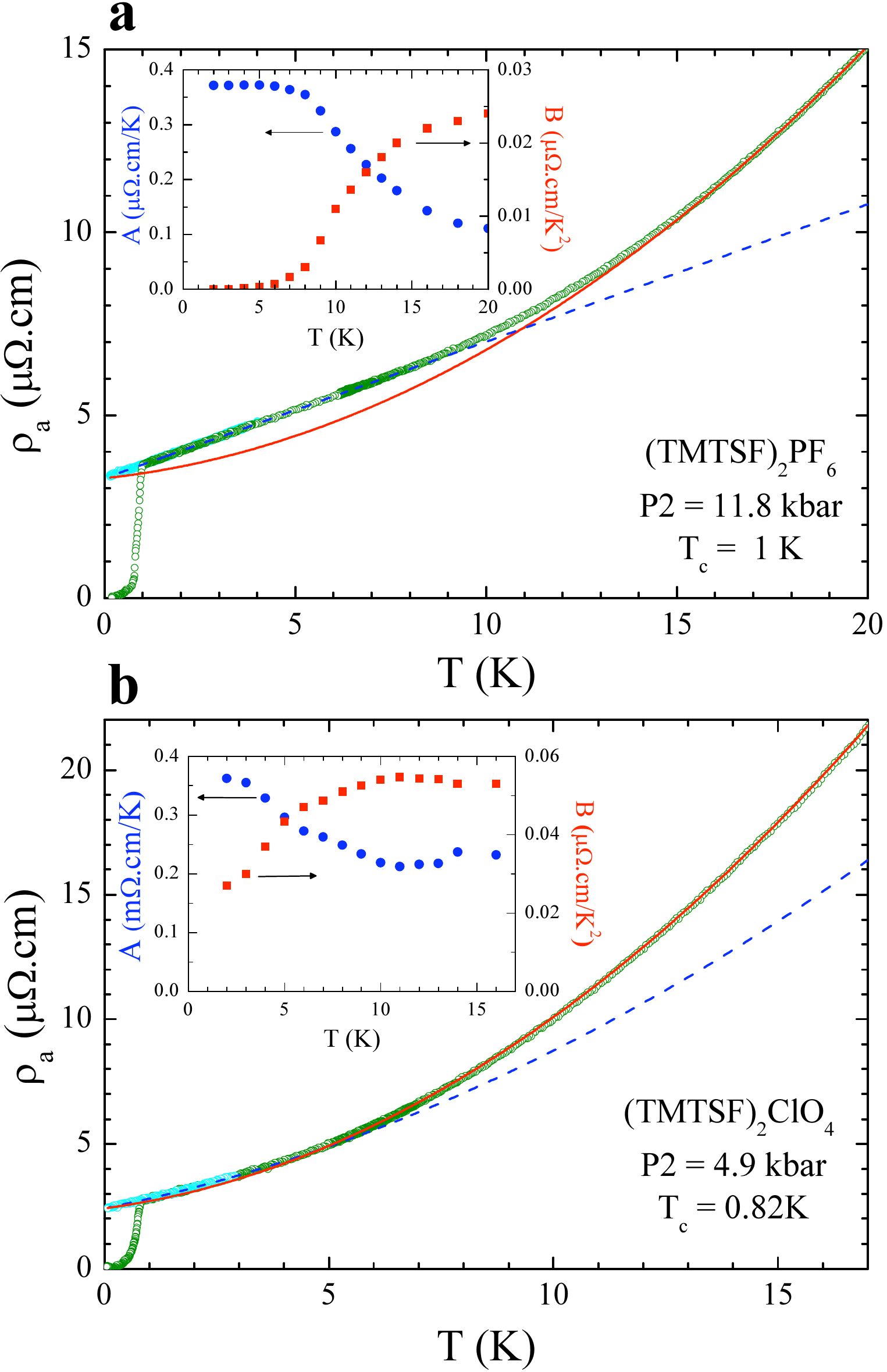}}
\caption{Typical temperature dependence of the longitudinal resistivity of \tmp6  below $20$ K (a), and \tmc at  below $17$ K (b), at zero field and under $H$ = 0.05 T along $c^*$ in order to suppress $SC$. The second order polynomial fit, $\rho(T) = \rho_0 + A(T)T + B(T)T^2$, according to the sliding fit procedure described in reference\cite{Doiron10} is shown for the  $T$ intervals $(2-6)$ K and  $(18-22)$ K or $(13-17)$ K in blue and red respectively. The top inserts provide  the  temperature dependence of the  $A$ and $B$ coefficients.}
\label{Fig3_RvsTABvsT} 
\end{figure*}

An extension of the transport analysis under pressure up to higher temperatures ($\approx 20$ K) in \tmp6\cite{Doiron10}   has suggested that the linear law is actually  the low temperature limit of a polynomial behaviour. 
Therefore, analysing  transport data with  a polynom, $\rho(T) = \rho_0 + AT + BT^2$ where $\rho_0$  is the linear extrapolation of the data below $4 K$ under a small magnetic field suppressing $SC$ and the  prefactors  $A$ and $B$ are determined by a sliding fit procedure\cite{Doiron10}. Both prefactors  are found to be $T$-dependent as displayed on fig(\ref{Fig3_RvsTABvsT}). 
\begin{figure*}[h]	
\centerline{ \includegraphics[width=1\hsize]{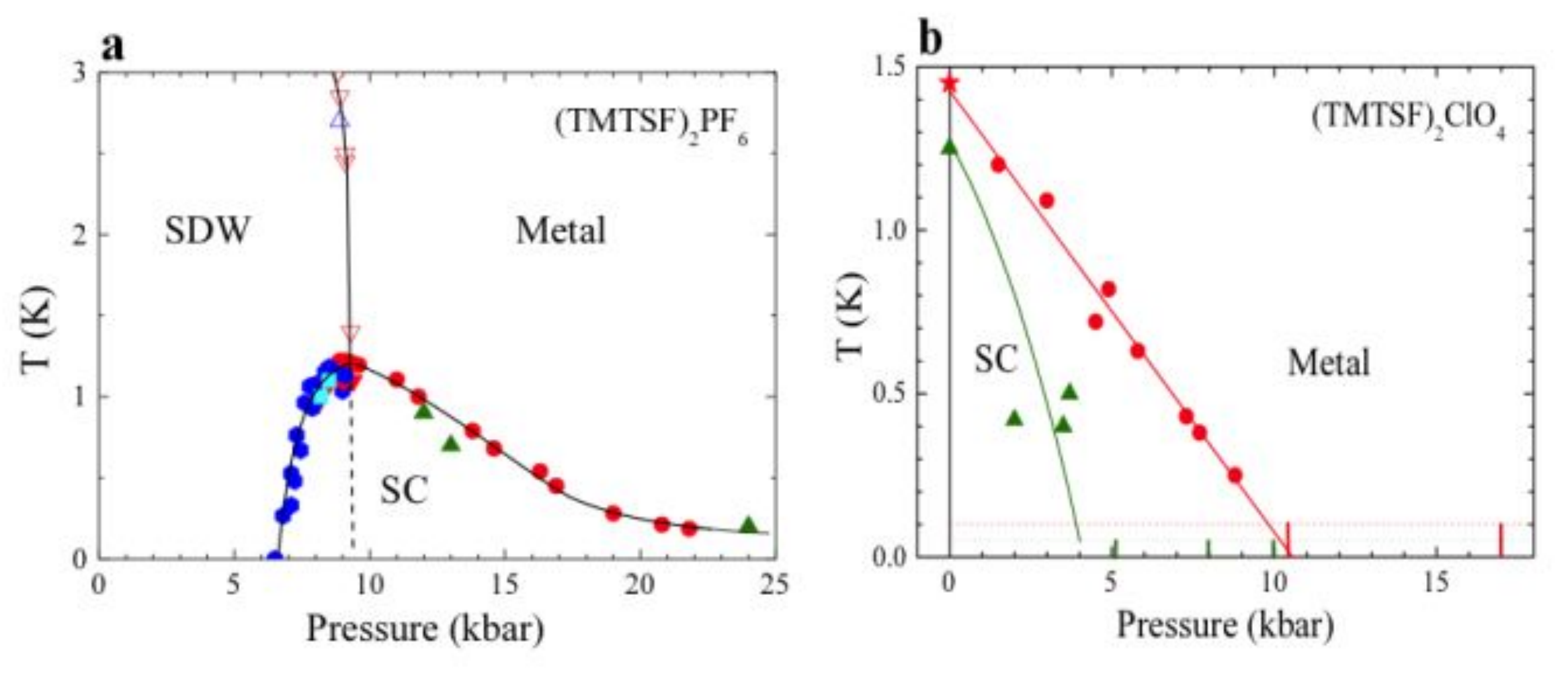}}
\caption{(a) ($P-T$) phase diagram of \tmp6 deduced from resistivity measurements. The data points below $9.4$ kbar (the coexistence regime) are deduced from measurements along the three crystallographic axes: down triangles for $\rho_a$, squares for $\rho_b$, hexagons for $\rho_c$, empty symbols for $T_{SDW}$ and full symbols for $T_{SC}$\cite{Vuletic02,Kang10}. Above the critical pressure ($9.4$ kbar), only longitudinal resistivity data are plotted: (red) circles from reference\cite{Doiron10} and (green) triangles from\cite{Ribault80}. (b) Pressure dependence of the superconducting transition of \tmc deduced from longitudinal resistivity measurements: (red) circles from reference\cite{Doiron10}  (the star at 1 bar is derived from a $\rho_c$ measurement\cite{Yonezawa08}) and (green) triangles from\cite{Parkin85}. Dashed-dotted horizontal lines (red or green) indicate the lowest reached temperature without superconductivity for both studies. The (red) continuous line is a linear fit of the data including the point at 1 bar.  \tc for \tmp6 although strongly decreasing under pressure does not reveal any critical pressure for the suppression of superconductivity \textit{at variance} with \tmc  in which no \tc can be detected at $10.4$ and $17$ kbar. Such a different behavior can be ascribed to the pair breaking effect of residual anion disorder in \tmc under pressure\cite{Joo05}, \textit{see text}.}
\label{DPhPFCl.pdf} 
\end{figure*}

There exists a high temperature domain ($T\geq$ 20 K) 
in which the regular $T^2$ electron-electron Umklapp scattering obeys a Kadowaki-Woods law\cite{Kadowaki86} and a low $T$ regime scattering   ($T\leq$8K) where the scattering behaves more and more   purely linear when pressure is close to \pc\cite{Doiron09}.
\begin{figure*}[h]	
\centerline{ \includegraphics[width=0.7\hsize]{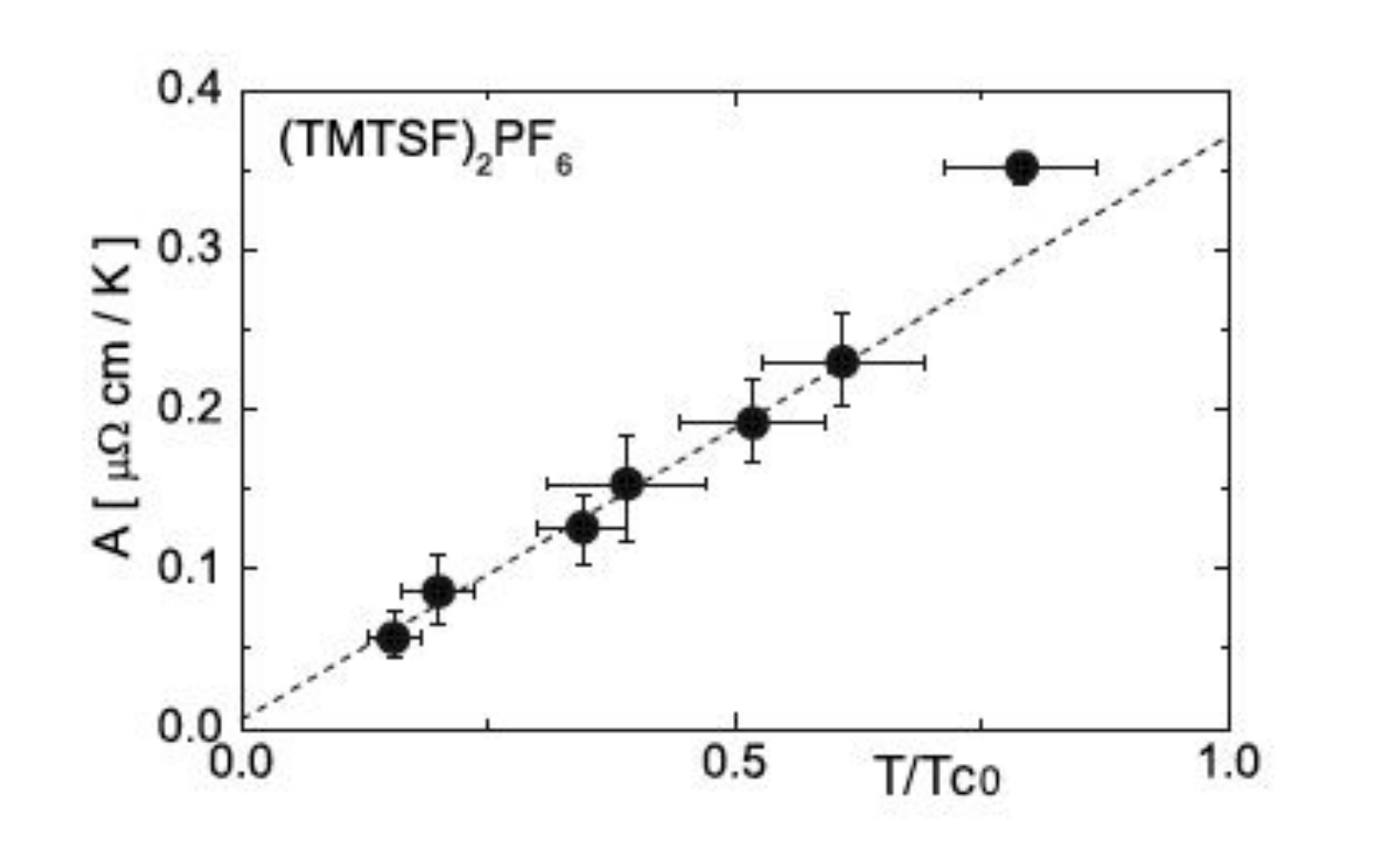}}
\caption{Coefficient $A$ of linear resistivity as a function of $T_c$ plotted versus $T_{c}/T_{c0}$  for \tmp6. $T_c$ is defined as the midpoint of the transition and the error bars come from the 10~\% and 90~\% points with  $T_{c0}=1.23K$ under the pressure of 8 kbar which provides the maximum \tc in the $SDW$/$SC$ coexistence regime.The dashed line is a linear fit to all data points except that at $T_c$~=~0.87~K,}
\label{AvsTcPF6.pdf} 
\end{figure*}

This  study has first established a  correlation between  the amplitude of the prefactor $A$ of the low temperature   linear dependence of the resistivity  and the value of the superconducting critical temperature \tc, fig(\ref{AvsTcPF6.pdf}). 

Superconductivity of \tmp6  has been  observed up to the maximum  pressure of this experiment (20.8 kbar) where $A$ is still  finite.   A similar relation between $A$ and \tc is obtained using the linear term of the $c^\star$ resistivity\cite{Auban11a}. As far as \tmc is concerned, the response of \tc to high pressure  is however different. A  behaviour  similar to that of \tmp6 is  noticed for both components of the resistivity  in the low pressure domain but the existence of a critical pressure where \tc vanishes is observed around 10.4 kbar.   Such a behaviour specific to \tmc can be attributed to the remaining weak disorder of the \C anions (even in slow cooled samples\cite{Pouget90}) suppressing \tc  when the mean distance between impurity centers and the superconducting coherence length become of the same order of magnitude as discussed in Sec(3.2). Since \tmp6 is not concerned by this  disorder effect it can be consider as an ultra pure type II superconductor.

\begin{figure*}[h]	
\centerline{ \includegraphics[width=0.7\hsize]{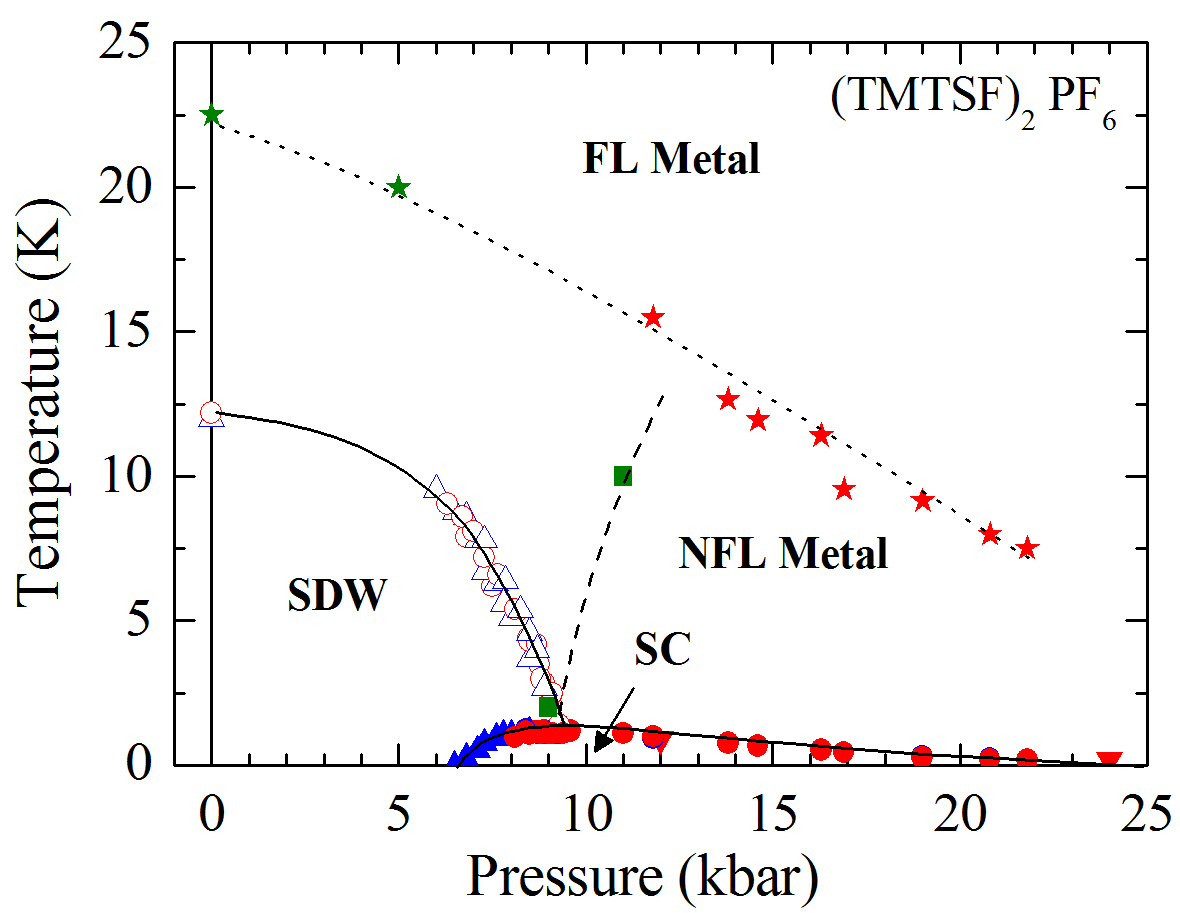}}
\caption{Phase diagram of \tmp6 showing the limit $T^\star$ for the domain of dominant AF fluctuations (NFL metal)  at the origin of the $T$ linear resistivity (dotted line). The two stars data points  above the $SDW$ ground state are given by the temperature for  the minimum of spin lattice $1/T_1$ relaxation \cite{Creuzet85}. In the FL metallic regime the temperature dependence of the resistivity is quadratic however high frequency optical conductivity data ($\omega \geq 1000cm^{-1}$)  still provide a power law decay which is  reminiscent of the Luttinger liquid behaviour\cite{Vescoli98,Pashkin06,Pashkin10}.  The dashed line shows the pressure dependence of the AF fluctuations energy scale ($\Theta$) derived from the temperature dependence of $T_{1}T$ using the data from reference\cite{Creuzet85}. A similar behaviour for $\Theta$ \textit{versus} $P$ has been obtained by Bourbonnais and Sedeki\cite{Bourbonnais11} using data of \tmp6 under pressure published by Brown and Wu\cite{Wu05,Brown08}. Be aware of the $\approx$ 3 kbar  difference   between pressure scales of UCLA and Orsay for the phase diagram of \tmp6. }
\label{DPh_PF6_Theta.jpg} 
\end{figure*}

 The  region of the $T-P$ phase diagram where antiferromagnetic fluctuations play a prominent role for the non elastic scattering can be conveniently defined by the temperature at which both contributions, the linear and the quadratic one are  equal, $T^\star=A/B$ where $A$ and $B$ are the  values of the parameters at every pressure taken in  the low and high  temperature limits respectively. Figure(\ref{DPh_PF6_Theta.jpg}) displays this strong fluctuations  domain. A similar behavior is obtained for \tmc\cite{Doiron10}.
 
The correlation between $A$ and \tc has suggested in turn a common origin for  the inelastic scattering of the metallic phase and pairing in the $SC$ phase \tmp6\cite{Doiron09}. 

 The decomposition of the inelastic scattering  term as the sum of linear and quadratic terms rather than a power law suggests that a regular Fermi liquid scattering channel is superimposed on a more unusual one, the latter being most likely connected to the scattering on low energy spin fluctuations.

This has been   shown indeed by scaling theory for the calculation of the electron-electron scattering rate close to  spin-density-wave ordering in a quasi-1D metal (the results are summarized in  reference\cite{Doiron10}). Near the critical pressure, where spin density wave connects with superconductivity (\textit{vide infra}), spin fluctuations are strong and their spectrum is sharply peaked at very low energy $\omega_{sf}$, which is comparable to or smaller than temperature $T$ (see, \textit{e.g.}, reference\cite{Bourbonnais09}). Under these conditions, their contribution yields a  clear linear  temperature dependence for the scattering rate, a known result for electrons interacting with low-energy  bosonic spin modes  in {\it two} dimensions (see e.g., \cite{Abanov03}). Moving away from critical pressure, spin fluctuations  decrease as shown on fig(\ref{DPh_PF6_Theta.jpg}), 

 This corresponds to an intermediate situation where electrons scatter on both low and sizable energy modes.  The former  modes are still responsible for  a linear term, though with a decreasing amplitude under pressure, while the latter modes favor  the opening of a  different scattering channel  at high energy which fulfills the Fermi liquid requirements ($\omega_{sf}\gg T$).   Scaling theory calculations  confirm that as one moves away from the critical pressure, the scattering rate is no longer perfectly linear in temperature above \tc,  but develops some curvature that is fitted quite satisfactorily by the $aT + bT^2$ form (see fig(10) of reference\cite{Doiron10}).

 \subsubsection{Scattering and pairing, a common origin}
Besides the seminal suggestion of Little for a non-phonon pairing mechanism in organics\cite{Little64}, several authors have have made proposals for non-phonon pairing in inorganic metals. Let us emphasize some of the early works. In the context of the  superconductivity in heavy fermions metals discovered the same year as  organic superconductivity\cite{Steglich79}, J. Hirsch has performed a Monte Carlo simulation of the Hubbard model showing an enhancement of anisotropic singlet pairing correlations due to the on site Coulomb repulsion leading eventually to an anisotropic singlet superconducting state\cite{Hirsch85}. One year later, L. Caron and C. Bourbonnais\cite{Bourbonnais86,Caron86} extending their theory for the generic \tm2x phase diagram to the metallic domain under pressure  made the first proposal of a gap equation for singlet superconductivity based on an interchain magnetic coupling. While the need for an interchain pairing had been noticed early by V. Emery and coworkers (see \cite{Emery83} and \cite{Beal86,Emery86}) the effective attraction in the gap equation of the Cooper channel is arising between carriers on neighbouring stacks as shown in the mean field renormalization theory\cite{Caron86,Bourbonnais88}. This attraction  is deriving from an interchain exchange interaction overcomingt the on-stack Coulomb repulsion. However, the renormalization treatment of Q-1-D conductors has received more recently a significant improvement, taking into account the interference between the diverging Cooper and Peierls channels\cite{Duprat01}.

It is recognized that the weak-coupling limit   explains fairly well the properties of the $SDW$ phases in \tsx materials both the   suppression of the $SDW$ phase under pressure and the stabilization of magnetic field-induced $SDW$ phases. The non interacting part of the quasi-one-dimensional electron gas model is defined in terms of a strongly anisotropic  electron spectrum yielding  an orthorhombic variant of the open Fermi surface in the $a-b$ plane of   the Bechgaard salts.  The spectrum $E({\bf k}) = v_F(|k|-k_F) -2t_\perp\cos k_\perp - 2t_\perp'\cos 2k_\perp$  as a function of the momentum ${\bf k}=(k,k_\perp)$  is characterized by an intrachain or longitudinal Fermi energy $E_F=v_Fk_F$  which revolves around 3000~K in (TMTSF)$_2$X \cite{Ducasse86,Lepevelen01}; here  $v_F$ and $k_F$ are the longitudinal Fermi velocity and wave vector. This energy is much larger than the interchain hopping integral $t_\perp$ ($\approx 200$K),  in turn much bigger than the second-nearest neighbor transverse hopping amplitude $t_\perp'$. The latter stands as the antinesting parameter of the spectrum which simulates the main influence of pressure in the model. 
The unnesting parameters of the band structure, $t^{'}_{b}$ and similarly $t^{'}_{c}$  for the $c^\star$ direction both play an important role  in the $T-P$ and $T-P-H$ phase diagrams of \tms2x. 

First, when $t^{'}_{b}$ exceeds a critical unnesting band integral of the order of   the $SDW$ transition temperature ($\approx 15-30K$ in case of complete nesting\cite{Montambaux86,Yamaji98}, the $SDW$ ground state is suppressed in favour of a metallic phase with the possibility of restoration of spin density phases under magnetic field along $c^{\star}$\cite{Ishiguro98}.
Second, the critical temperature for the stabilisation of the $SDW$ subphases, $T_{FISDW}(H)$ should be steadily  increasing from zero in a 2D conductor or in a fully nested 3D conductor in the "standard model"\cite{Heritier84,Heritier86}. However, since the real system is neither 2D nor perfectly nested ( $t^{'}_{c}$$> $0), there exists a threshold field $H_{T}$ for the appearance of $SDW$ subphases defined by $T_{FISDW}(H_{T})$=$t^{'}_{c}$\cite{Montambaux86}.  Experimentally the threshold field is of the order of $2 T$\cite{Hannahs89,Cooper89}

Second,  within the framework of a weak-coupling limit  the problem of the interplay between antiferromagnetism and superconductivity in the Bechgaard salts has been worked out using the renormalization group (RG) approach \cite{Bourbonnais09,Nickel05} as sketched below taking into account only the 2D problem.  
 The RG integration of high energy electronic degrees  of freedom was  carried out  down to the Fermi level, and leads to a  renormalization of the couplings at  the temperature $T$ \cite{Duprat01,Nickel06,Bourbonnais09}.  The RG flow   superimposes  the $2k_F$ electron-hole (density-wave) and Cooper pairing many-body processes which combine and interfere    at every  order of perturbation. As a function of the `pressure' parameter  $t_\perp'$, a   singularity  in the scattering amplitudes signals an instability of the metallic state  toward the formation of an ordered state at some characteristic temperature scale. At low $t_\perp'$, nesting is sufficiently strong to induce a $SDW$ instability in  the temperature range  of experimental  $T_{\rm SDW}\sim 10-20$~K. When the antinesting parameter approaches the threshold $t_\perp'^*$  from below  ($t_\perp'^* \approx 25.4~{\rm K}$, using the above  parameters), $T_{\rm SDW}$ sharply decreases and  as a result of interference, $SDW$ correlations ensure  Cooper pairing attraction in the superconducting $d$-wave ($SCd$) channel. This   gives rise to  an    instability  of the normal state   for the onset of  $SCd$ order at  the temperature $T_c$ with pairing   coming from antiferromagnetic spin fluctuations between carriers of neighbouring chains. Such a pairing model is actually supporting the conjecture of interchain pairing in order for the electrons to avoid the Coulomb repulsion made  by V. Emery in 1983 and 86\cite{Emery83,Emery86}.
 
 The calculated phase diagram with reasonable parameters taking   $g_1=g_2/2 \approx 0.32$ for  the backward and forward scattering amplitudes  respectively and $g_3\approx 0.02$ for the  longitudinal  Umklapp scattering  term captures the essential features of the experimentally-determined phase diagram of\\ \tmp6~\cite{Bourbonnais09,Bourbonnais11} to be compared with the experimental diagram on fig(\ref{DPh_PF6_Theta.jpg}). 
 
Third, Sedeki and Bourbonnais\cite{Sedeki10} have proceeded to an  evaluation of the imaginary part of the one-particle self-energy.  In addition to the regular Fermi-liquid component which goes as $T^2$ low frequency spin fluctuations  yield $\tau^{-1} = aT\xi $, where $a$ is constant and the antiferromagnetic correlation length $\xi(T)$ increases according to $\xi = c(T + \Theta)^{-1/2}$ as $T \rightarrow T_c$, where $\Theta$ is the temperature scale for spin fluctuations \cite{Sedeki10}.  It is then natural to expect the Umklapp resistivity to contain  (in the limit $T \ll \Theta$) a linear term $AT$ besides the regular $BT^2$, whose magnitude would presumably be correlated with $T_c$, as both scattering and pairing are caused by the same antiferromagnetic correlations. The observation of a $T$-linear law for the resistivity up to 8 K  in \tmp6 under a pressure of 11.8 kbar as displayed on fig(\ref{Fig3_RvsTABvsT})   is therefore  consistent with the value of $\Theta=8K$ determined from NMR relaxation at 11 kbar on fig(\ref{T1TvsT_PF6.jpg}).   
 \subsubsection{$SDW$ conductive fluctuations very close to \tc}
However, these recent studies of the low temperature regime were still unable to explain the very particular sublinear behaviour of the resistivity  observed  up to about three or four times \tc when the pressure is located in the very vicinity of the critical pressure \pc suppressing the insulating spin density wave ($SDW$) phase in \tmp6\cite{Kang10}. Therefore, in the vicinity of \pc, two precursor regimes can be distinguished, first the $SC$ transition\textit{ per se} which is usually rather narrow in temperature, ($\Delta T \approx 0.1-0.2 K)$ and second, a broader temperature regime above \tc up to about 4K in which the sublinear resistivity is observed.  

These two regimes have been reexamined using transport along the transverse $c$ direction.
\begin{figure*}[htbp]	
 \centerline{\includegraphics[width=0.5\hsize]{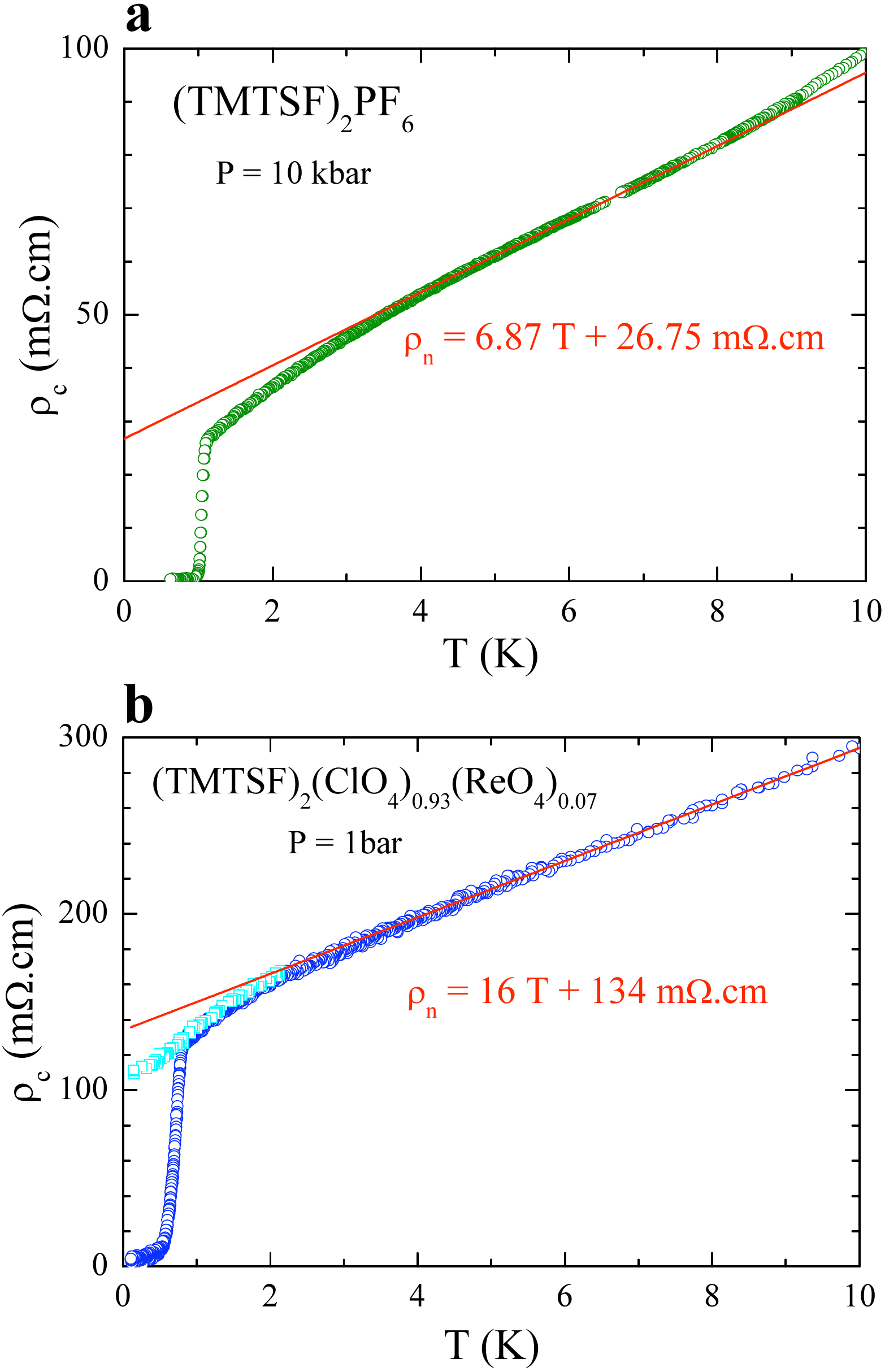}}
\caption{Temperature dependence of $\rho_c$ for \tmp6 at a pressure close to 10 kbar, at zero field (a) and  \tmx\, $\mathrm{x}=7\%$ at ambient pressure at zero field and under a small magnetic field applied along $c^{\star}$ in order to suppress $SC$ (b) . The line is the linear fit for the normal state resistivity $\rho_n$ between 4 and 8 K.}
\label{SDWFluct_PF6_ClO4} 
\end{figure*}

The major problem for the evaluation of an excess conduction coming from a collective motion independently of its exact origin  is the proper determination of the temperature dependence of  the background conduction due to the quasi particle scattering. Fortunately, such an investigation,  carried out recently on \tmp6 and \tmc and summarized in the previous Section  has led to the determination of this background resistivity with a  scattering rate behaving linearly in temperature (although not too close in temperature to \tc) and tightly connected to the value of \tc (which can be varied by pressure)\cite{Doiron09,Doiron10}.

 Two cases for the approach to superconductivity of \tsx are displayed in fig(\ref{SDWFluct_PF6_ClO4}) when the  compounds are located in their phase diagram close to the critical pressure for the stabilisation of $SC$. This pressure is real for \tmp6 and amounts to 9.4 kbar\cite{Vuletic02,Kang10}. In \tmc, it is more likely a \textit{virtual}  negative pressure since $SC$ is already stable under ambient pressure.

The first case  shown on fig(\ref{SDWFluct_PF6_ClO4}) is  \tmp6  under a  pressure around 10 kbar.
The second case refers to \tmc at ambient pressure in which \tc has been slightly depressed down to 0.75 K via the pair breaking of a controlled amount of non magnetic defects\cite{Joo05}, see Sec (3.2).

We  notice that in a material such as \tmp6 where superconductivity arises in the neighborhood of a $SDW$ ground state stable below the critical pressure \pc we can anticipate for the metallic phase above \pc a remnence of fluctuating $SDW$'s  at low temperature. Actually, these low frequency AF fluctuations ($q=2k_F$) are well known to contribute to the nuclear spin-lattice relaxation below 20 K or so\cite{Bourbonnais08}. Accordingly, they may be able to provide  additional conductivity very much like incommensurate fluctuating $CDW$'s contribute in a low dimensional metal above  the Peierls transition\cite{Jerome82} given the incommensurate character of the $SDW$ firmly established by the observation of a double horn shape of the $^{13}C$ NMR spectrum in \tmp6\cite{Barthel93}.
\begin{figure*}[tb]
 \centerline{\includegraphics[width=0.6\hsize]{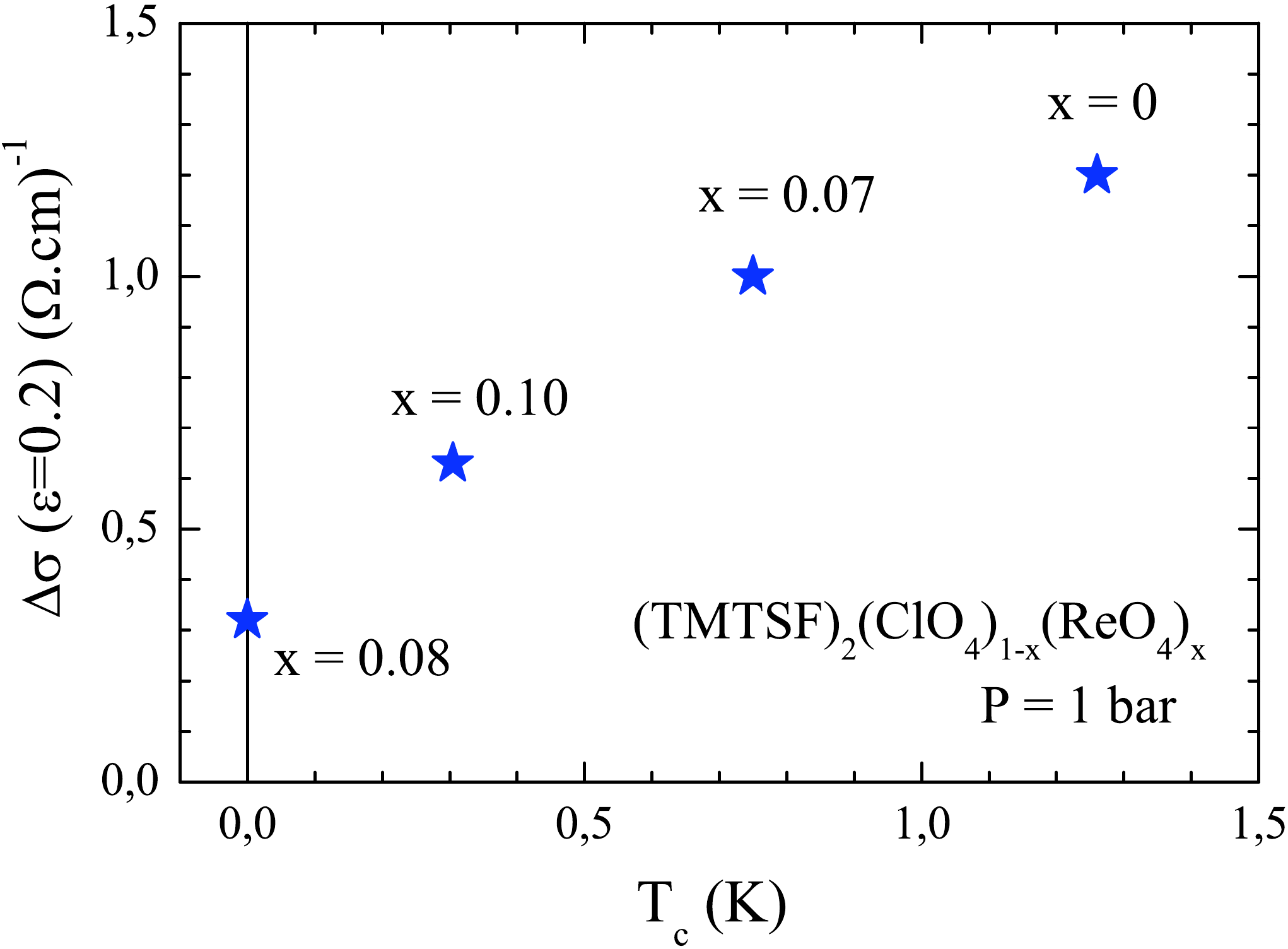}}
\caption{Paraconductive contribution characterised  by $\Delta\sigma $ ($\epsilon$ =$(T-T_c)/T_c= 0.2)$ (star symbols) in the precursor regime, versus $T_c$  in the alloy series of \tmx samples at 1 bar.The  sample x=0.1 is superconducting whereas the one  x=0.08,  is not superconducting above 0.1 K although still displaying some fluctuating $SDW$ conduction. This is nothing more that the indication that the nominal \R concentration is not the best way to characterize the amount of defects, at least for large concentrations. }
\label{fig4_epl}
\end{figure*}
The depression of the resistivity below 3 K is insensitive to the application of a magnetic field even up  to 5 T \textit{albeit} strongly suppressed by non magnetic defects. This feature enables to rule out any superconducting origin.  It becomes then quite sensible  to attribute it to the sliding of $SDW$ fluctuations contributing to the paraconductivity $\Delta\sigma $.
 We can ascribe the drop of $SDW$ paraconductivity displayed on fig(\ref{fig4_epl}) in  doped samples to the net effect of pinning  centers. 

Substituting  $\mathrm{ReO_4}$  to $\mathrm{ClO_4}$ amounts to a decrease of  the mean distance between impurities. When the  mean distance becomes  comparable to the $SDW$ coherence length, the impurity pinning suppresses $\Delta\sigma$ heavily\cite{Jerome82}.
Data for the  $SDW$ coherence length support this explanation since the zero-temperature longitudinal length $\xi_{0a}$ is of order of 32 nm in \tmp6\cite{Gruner94}\textit{ i.e}, 44 times the distance between \cl \,anions. Consequently, it  is reasonable that the $SDW$ is pinned by defects when the mean distance between pinning centers  $\approx$7.3 nm  (for 10$\%$ doping)  is four times smaller than the coherence length leading  in turn to a decrease  of the fluctuation conduction independently of the current direction.
The existence of a non perfect $\mathrm{ClO_4}$ anion ordering in pure \tmc may also be  the origin for a   $SDW$ paraconductivitng conribution appearing to be  smaller  in \tmc than in \tmp6.

\section{Summaryzing, magnetism everywhere}
\label{sec:5}
\tmp6 - \textit{Most remarkable electronic material ever discovered}. If Paul Chaikin says so on his personal Web site\cite{Chaikinweb}, it must be true because Paul is rarely wrong with his statements.

These superconductors are starting with materials where correlations and Umklapp scattering render the low dimensional electron system truly one dimensional as displayed by the 1D Mott localization on the left part of the generic diagram, fig(\ref{generic2.pdf}). We may call this regime the strong coupling limit of the diagram. As pressure enhances the bare  interstack coupling, 1D Mott localization is quickly suppressed when the Mott gap and the transverse coupling are of the same order. Then begins the weak coupling regime of the phase diagram with a Fermi surface whose nesting is responsible for the onset of a weak itinerant antiferromagnet at low temperature (the $SDW$ state) leading in turn to an insulator since nesting of the Fermi surface is nearly perfect. Increasing the transverse coupling still further under pressure suppresses the $SDW$ phase to the benefit of a metallic state becoming superconducting at low temperature.

However, superconductivity is not emerging from a Fermi liquid.  NMR  experiments reveal the existence of AF fluctuations dominant in the hyperfine $1/T_1$ at low temperature. Measurement of the spin susceptibility via relaxation  data also reveal the 2D nature of these fluctuations. 2D AF fluctuations are actually governing  the low temperature transport and the onset of superconductivity, both being tightly connected. 

\begin{figure*}[htbp]	
 \centerline{\includegraphics[width=0.7\hsize]{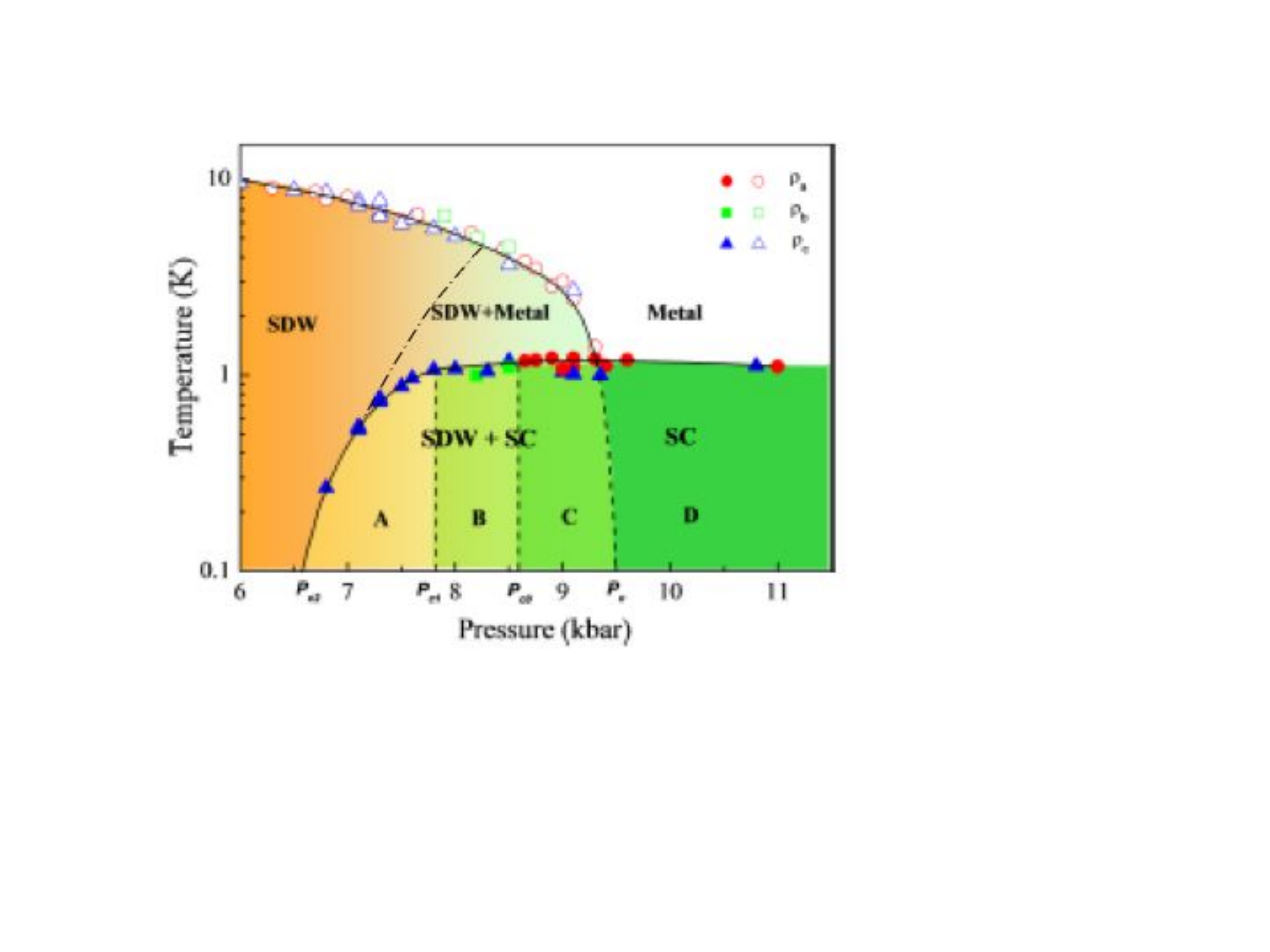}}
\caption{ Detailed determination of the phase diagram of \tmp6 under zero magnetic field from resistivity measurements in the coexistence domain, according to refs\cite{Kang10,Vuletic02}. It is hard to tell where $P_{QCP}$ could be located in this diagram but possibly in the region C of the figure. } 
\label{coexistence}
\end{figure*}

The weak coupling RG theory of the Q1D electron gas developed by Bourbonnais and his colleagues does reproduce very nicely most experimental features.  Interestingly, they came to the conclusion that the two types of order are intimately connected namely, the non trivial result  that antiferromagnetism and superconductivity love each other!
Cooper pairing reinforces $SDW$ correlations.  AF fluctuations are also responsible for the linear-$T$  resistivity which should increase as pressure approaches a magnetic quantum critical point. Actually the QCP is probably unaccessible for \tmp6 since it may correspond in terms of pressure  to the zero temperature extrapolation of the $SDW$ transition line.
Hence $P_{QCP}$ is likely to be smaller than $P_c$ hidden inside the $SDW$/Metal($SC$) coexistence regime\cite{Vuletic02,Kang10}, fig(\ref{coexistence}).
A very detailed reinvestigation of the coexistence regime   under pressure should be quite a worthwhile experimental program. For such a study helium gas would be a particularly well adapted pressure medium  since this could allow a very fine control of the pressure without warming the whole pressure cell up to room temperature between each runs. Admittedly, the study of Seebeck and Nernst coefficient in \tmp6  under pressure should be very rewarding experiments\cite{Kim11,Kim10}.  For \tmc the QCP is likely to be located at a negative pressure. This experiment is obviously impossible but an uniaxial elongation experiment has shown that an insulating ground state can be reactived in the relaxed state of \tmc when some elongation is applied along the $b$ direction\cite{Kowada07}  leading to a  decrease of  the $b$ interchain coupling. When this coupling  becomes smaller than the critical value  stabilizing the metallic phase a $SDW$ ground state is likely to be stabilized.

\begin{figure*}[htbp]	
 \centerline{\includegraphics[width=0.7\hsize]{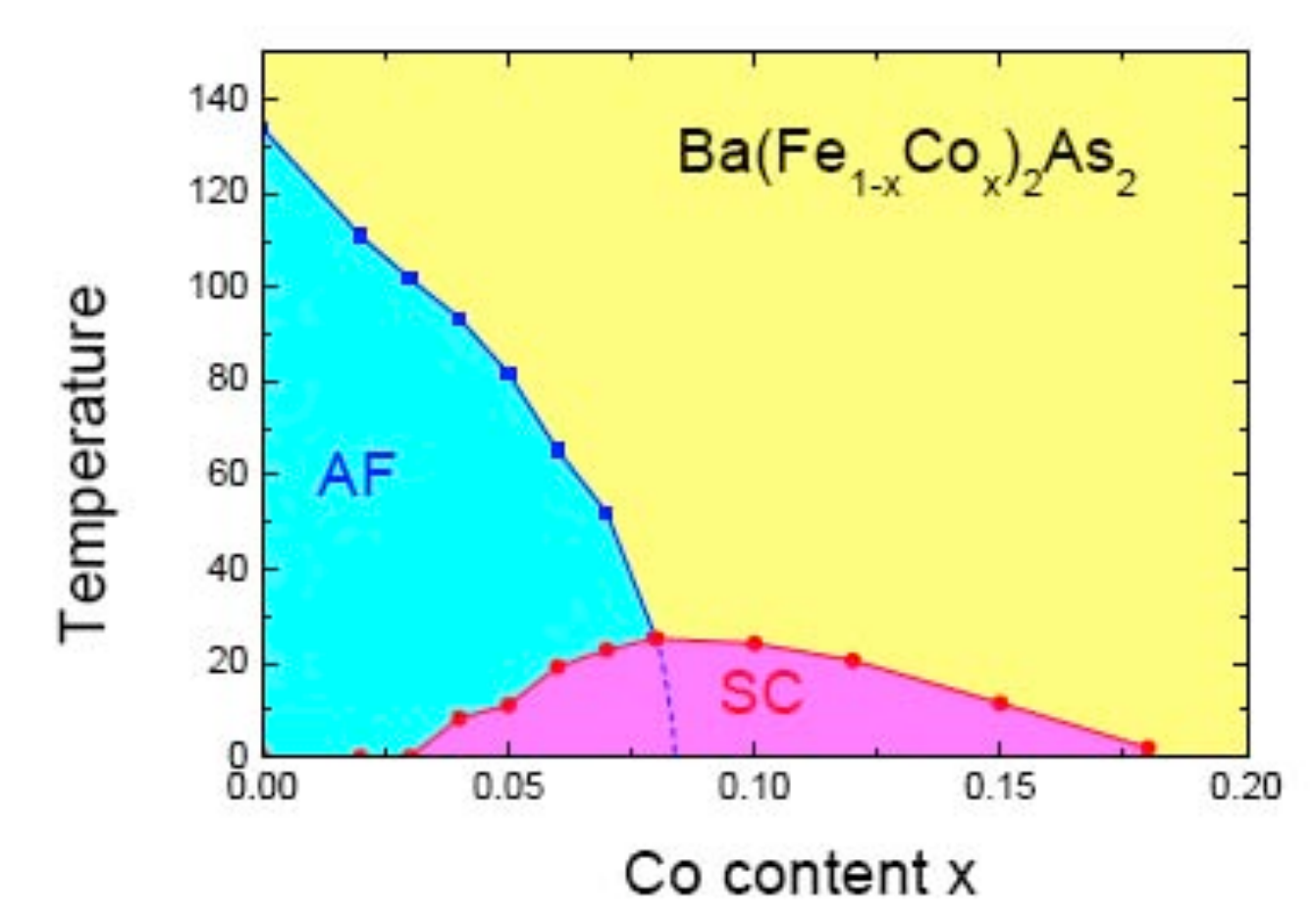}}
\caption{Phase diagram of Ba(Fe$_{1-x}$Co$_x$)$_2$As$_2$ pnictides according to Fang et-al\cite{Fang09}. Notice the close resemblence with fig(\ref{DPh_PF6_Theta.jpg}). } 
\label{pnictidediagram.pdf}
\end{figure*}

 It is worth noting that the non Fermi liquid properties of the metallic background of Bechgaard salts prior to the onset of superconductivity may be quite a general behaviour when  magnetism is close to superconductivity as noticed in reference\cite{Doiron09}, see also \cite{Taillefer10} for a review. For instance, the phase diagram  of the iron-pnictide superconductors Ba(Fe$_{1-x}$Co$_x$)$_2$As$_2$,  with its adjacent semi-metallic $SDW$ and superconducting phases\cite{Fang09,Chu09},  presents a striking family resemblence   with the diagram of  \tm2x.  Comparison with the resistivity data of Fang \textit{et-al}~\cite{Fang09} and Chu \textit{et-al}~\cite{Chu09} on the pnictide superconductor Ba(Fe$_{1-x}$Co$_x$)$_2$As$_2$ suggests that the findings on \PF and \tmc~may be a more general property of metals near a $SDW$ instability. The phase diagram of Ba(Fe$_{1-x}$Co$_x$)$_2$As$_2$ \cite{Chu09,Fang09},  is  similar to that of \tmp6, with $T_{SDW}$~and \tc both enhanced by a factor of about 20, and just above the critical doping where $T_{SDW}$~$\rightarrow$~0 (at $x \approx$~0.08), its resistivity is purely linear below ~125 K, down to at least \tc$ \approx$~25 K, see fig(\ref{pnictidediagram.pdf}). 
  Furthermore,  in the context of high-\tc cuprates, it has long been known that the low temperature resistivity of strongly-overdoped non-superconducting samples has the form $\rho(T) = \rho_0 + BT^2$, as in Tl$_2$Ba$_2$CuO$_{6+\delta}$ (Tl-2201) at $p = 0.27$ \cite{Manako92} and La$_{2-x}$Sr$_x$CuO$_4$ (LSCO) at $p = 0.33$ \cite{Nakamae03}. It was also shown that the evolution of $\rho(T)$ from  $\rho(T) = \rho_0 + AT$ near optimal doping to $\rho(T) = \rho_0 + BT^2$ at high doping is best described by the approximate form  $\rho(T) = \rho_0 + AT+ BT^2$ at intermediate doping \cite{Mackenzie96,Proust02}.

  Recently, a linear resistivity at low temperature has been observed in high-field resistivity data on overdopedÊ
Ê ÊNd-LSCO\cite{Daou09} and LSCO \cite{Cooper09} and it was emphasized that the $A$ coefficient  in hole-doped cuprates also correlates with\,Ê\tc,Êvanishing close to the point where superconductivity disappears\cite{Daou09,Cooper09,Taillefer10}.
Electron-doped cuprates provide an other situation where a $T$-linear resistivity and the proximity of an antiferromagnetic regime prevail in their phase diagram when Ce doping is varied. 
The resemblence between  $\mathrm{La_{2-x}Ce_{x}CuO_{4}}$ and \tmp6 has been acknowledged recently\cite{Jin11} supporting earlier observation made on $\mathrm{Pb_{2-x}Ce_{x}CuO_{4}}$\cite{Fournier98}. 
However, in the case of electron-doped cuprates there still remains some uncertainty regarding the exact location of the QCP in their phase diagram\cite{Jin11}. 

\section{Conclusion and organic superconductivity in 2012}

 I have tried to show in this short  presentation  why  the  research on organic superconductors has been initiated and why it  has started in the seventies  in the context of a search for high temperature superconductivity    boosted by two stimulating results. 
 
 First, a  model suggested in 1964 by W. A. Little based on a phonon-less electron pairing leading   to high \tc organic superconductivity.  It is worth citing V.L.Ginzburg \cite{Ginzburg89}: \textit{I believe that it was undoubtedly the discussion of the possible exciton mechanism of superconductivity that stimulated the search for  such superconductors and studies of them}. Ginzburg was thinking at the Little's proposal and the  organic superconductors and intercalated layered superconductors which came later. 

Second, the claim made by the Penn group in 1973 for precursor signs of a phonon-mediated superconductivity in the first organic conductor TTF-TCNQ  just above the onset of a one dimensional metal insulator Peierls transition in  \tq,  the first  organic metal. 
 
The model of Little and the claims of the Penn group were probably overemphasized, but boosted the whole field at its start. The discovery of   superconductivity in the organic salt \tmp6 in  1980 has shown how interesting  this class of 1D organic conductors materials can be for the study  of the physics of low dimensional conductors.

Research on organic superconductors  has certainly  been somehow overshadowed by  the equally   fascinating oxides and pnictides superconductors exhibiting supercondivity at much higher \tc but  Ginzburg had quite a pragmatic approach to organics when he  wrote again : \textit{the organic superconductors are clearly interesting by themselves or, to be more  precise, irrespective of the high \tc problem }. 

 High pressure has played a   determining role for this discovery as often in the field of superconductors and governs the variety of phenomena which can be studied in a single sample keeping both the structure and the chemical purity constant. Furthermore, pressure is  governing their physical properties, 1D Mott and Wigner localization and the wealth of various  instabilities at low temperature which have shown how organic conductors compare fairly well  in terms of scientific interest with the celebrated class of high \tc cuprates. In particular,the interplay between the  Mott-driven localization of the 1D correlated electron gas and the 2D deconfinement under pressure is    an issue central to  the dimensionality crossover and also  how the restoration of  a Fermi liquid is achieved in quasi-one-dimensional conductor.

 In such a  brief account I could not go  through the detailed  physical properties of the Bechgaard and Fabre salts series of organic superconductors but  the generic phase  diagram that was gradually built over the years around these two series of compounds under either hydrostatic or  chemical pressure stands out as a model of unity for the physics of low dimensional correlated systems. Much effort went to  explain  the  multifaceted phase diagram of \TM\ as a whole, an attempt that also proved to be an active quest of unity for the theory. 

At the  starting point  of the study of the Bechgaard salts more than thirty years ago, superconductivity was certainly  one of the hardest part of  the phase diagram to both explain and  characterize. The  mechanism of organic superconductivity in quasi-one-dimensional molecular crystals has been a related  key issue in want of a satisfactory explanation for several decades.  The statement of  P.W. Anderson when he  was interviewed a few years ago  in a Physics Today issue  dedicated to superconductivity,  mentioning organic superconductors was: \textit{These are still a complete mystery}. Hopefully,  this   has   probably to be revised in the light of 
the recent experimental advances which   confirm  the non conventional nature of superconductivity in \TM, as shown by the   symmetry of the superconducting order parameter, as well as the  issue of the presence and location of nodes for the gap which have now emerged from the experiments. 

From a theoretical view point it is clear that electron correlations have walked in the superconducting pairing problem. The extensive experimental evidence  in favor of  the  systematic emergence of superconductivity in \TM \ just below their  stability threshold for antiferromagnetism  has shown the need for a unified description of electronic excitations that lies at the core of  both density-wave and superconducting correlations. In this matter, the  recent progresses achieved by the renormalization group method for the 1D-2D electron  gas model  have resulted  in definite predictions about the possible symmetries of the superconducting order parameter when a purely electronic mechanism is involved,  predictions that often differ from  phenomenologically based   approaches to superconductivity but are in fair agreement with the recent experimental findings.  Although one can never be hundred percent sure at the moment, as far as recent results on 1D organic superconductivity are concerned  magnetism and superconductivity cooperate.
 
What is emerging from   the past work on these prototype 1D  organic superconductors is   their  very simple electronic nature  with only a single band at Fermi level, no prominent spin orbit coupling and extremely high chemical purity and stability. They should be considered in several respects as model systems to inspire the physics of the more complex high \tc superconductors, especially for  pnictides and electron-doped cuprates. Most concepts discovered in these simple low dimensional conductors should be very useful for the study of other low dimensional systems such as carbon nanotubes or artificial 1D structures, \textit{etc},.....The pairing mechanism behind organic superconductivity is likely different from the proposal made by Little but it is nevertheless a phonon-less mechanism, at least in \TM \,superconductors .

Of course, such a short overview is missing    several  of the very important aspects of the physics of 1D conductors namely,  the  magnetic field
confinement discovered in the 1D's leading to the phenomenon of field induced spin density wave phases\cite{Ribault83,Chaikin83,Gorkov84} and  the quantization of the Hall effect in these phases\cite{Cooper89,Hannahs89,Heritier84}. Furthermore, the  angular dependent magnetoresistance  specific to these anisotropic conductors \cite{Osada91,Naughton91}, (see \cite{Osada06} for a recent survey), the so-called Lebed-Osada-Danner-Chaikin oscillations in \tms2x provide a nice illustration for the new features of the quasi-1D conductors\cite{Lebed86a}.  

Given the    experimental constraints and difficulties tied to the use of extremely low temperature and high  pressure conditions  in  \TM, their properties will certainly continue to attract major experimental efforts in the next few years. 

Let me tell the coming generation,
\textit{ there is still plenty of food  for thought in  research on organic superconductivity}!
 
 \section{Acknowledgments}
 
 The search, discovery and study of organic superconductors has been a major research domain for the Solid State Physics Laboratory at Orsay over the last thirty years. We are all indebted to Professor Jacques Friedel for his contributions to this research at the Laboratoire de Physique des Solides. 
 
 I have been offered in 1967 by Jacques Friedel the  creation of  the group specialized in the study of metals and alloys under high pressure and low temperature.  My first objective, the search for the excitonic instability in ytterbium  under pressure and low temperature had not been a great success but mastering  high hydrostatic pressure experiments turned out to be a decisive asset for the investigation of periodic lattice distorted layer conductors and most definitely   in the field of organic conductors and superconductors.  
 
 Professor Friedel has  always supported and encouraged us with his wise advices and clever suggestions.  This period has also been   remarkably fruitful for the cooperation between chemists, theoreticians and experimentalists. I wish to address my sincere and deepest thanks to all my co-workers not only at Orsay but also at Sherbrooke (Qu\'ebec), Copenhagen and Kyoto. I wish to address my warm thanks to Professor Yonezawa who has contributed to  fig(\ref{volovik-effect_for_Review_ProfJerome_.pdf}) deriving from his recent illuminating results\cite{Yonezawa12}.

\end{document}